\documentclass{emulateapj2}

\newcommand{\msun}{M$_{\sun}$}
\newcommand{\ldl}{$\lambda/{\Delta}{\lambda}$}

\newcommand{\teff}{T$_{eff}$}
\newcommand{\lbol}{$\log{{\rm L}_{bol}/{\rm L}_{\sun}}$}
\newcommand{\ki}{\ion{K}{1}}

\newcommand{\nai}{\ion{Na}{1}}

\newcommand{\lii}{Li~I}

\newcommand{\ammon}{NH$_3$}
\newcommand{\meth}{CH$_4$}
\newcommand{\wat}{H$_2$O}

\newcommand{\kms}{km~s$^{-1}$}

\newcommand{\vtan}{$V_{tan}$}

\newcommand{\namea}{2MASS~J08503593+1057156}
\newcommand{\namesha}{2MASS~J0850+1057}
\newcommand{\nameb}{2MASS~J17281150+3948593}
\newcommand{\nameshb}{2MASS~J1728+3948}

\slugcomment{Submitted to AJ on 23 August 2010; Accepted for publication 5 October 2010}

\shorttitle{HST Brown Dwarf Multiples}
\shortauthors{Burgasser, Bardalez Gagliuffi \& Gizis}

\begin{document}

\title{Hubble Space Telescope Imaging and Spectral Analysis of Two Brown Dwarf Binaries at the L Dwarf/T Dwarf Transition\footnote{Based on observations made with the NASA/ESA {\em Hubble Space Telescope}, obtained from the Data Archive at the Space Telescope Science Institute, which is operated by the Association of Universities for Research in Astronomy, Inc., under NASA contract NAS 5-26555. These observations are associated with program GO-9843.}}

\author{Adam J.\ Burgasser\altaffilmark{1,2}}
\affil{Center for Astrophysics and Space Science, University of California San Diego, La Jolla, CA 92093, USA; aburgasser@ucsd.edu}
\author{Daniella C. Bardalez Gagliuffi\altaffilmark{1}}
\affil{Massachusetts Institute of Technology, Kavli Institute for Astrophysics and Space Research, 77 Massachusetts Avenue, Cambridge, MA 02139, USA}
\and
\author{John E.\ Gizis}
\affil{Department of Physics and Astronomy, University of Delaware, Newark, DE 19716, USA}

\altaffiltext{1}{Visiting Astronomer at the Infrared Telescope Facility, which is operated by the University of Hawaii under Cooperative Agreement no. NCC 5-538 with the National Aeronautics and Space Administration, Science Mission Directorate, Planetary Astronomy Program.}
\altaffiltext{2}{Hellman Fellow.}

\begin{abstract}
We present a detailed examination of the brown dwarf multiples
2MASS~J08503593+1057156 and 2MASS~J17281150+3948593, both suspected of
harboring components that straddle the L dwarf/T dwarf transition.  
Resolved photometry from {\em Hubble Space Telescope}/NICMOS show
opposite trends in the relative colors of the components, with the secondary of 2MASS~J0850+1057 being redder than its primary, while that of 2MASS~J1728+3948 is bluer.  We determine near-infrared component types by matching combined-light, near-infrared spectral data to binary templates, with component spectra scaled to resolved NICMOS and $K_p$ photometry.
Combinations of L7 + L6 for 2MASS~J0850+1057 and L5 + L6.5 for 2MASS~J1728+3948 are inferred.  Remarkably, the primary of 2MASS~J0850+1057 appears to have a later-type classification compared to its secondary, despite being 0.8--1.2~mag brighter in the near-infrared, while the primary of 2MASS~J1728+3948 is unusually early for its combined-light optical classification.  Comparison to absolute magnitude/spectral type trends also distinguishes these components, with 2MASS~J0850+1057A being $\approx$1~mag brighter and 2MASS~J1728+3948A $\approx$0.5~mag fainter than equivalently-classified field counterparts.  We deduce that thick condensate clouds are likely responsible for the unusual properties of 2MASS~J1728+3948A, while 2MASS~J0850+1057A is either an inflated young brown dwarf or a tight unresolved binary, making it potentially part of a wide, low-mass, hierarchical quintuple system.
\end{abstract}


\section{Introduction}

The L dwarfs and the T dwarfs are the two lowest luminosity spectral classes of very low-mass stars and brown dwarfs known today, encompassing effective temperatures ({\teff}) from $\sim$2200~K down to $\sim$600~K (\citealt{2004AJ....127.3516G, 2008MNRAS.391..320B, 2009ApJ...702..154S}; see also \citealt{2005ARA&A..43..195K} and references therein).
Spectroscopic studies of these sources reveal atmospheres that are remarkably diverse. L dwarfs having abundant molecular gas species 
and clouds of condensates in their photospheres, while T dwarfs
have relatively cloud-free photospheres and more complex molecular gas species including
{\meth}, {\ammon} and CO$_2$ \citep{1995Sci...270.1478O, 2004ApJS..154..418R, 2010arXiv1008.3732Y}.  Condensate cloud properties are  believed responsible for near-infrared spectral and color variations among equivalently-classified L dwarfs \citep{2004AJ....127.3553K, 2008ApJ...674..451B, 2008ApJ...686..528L}, temporal variability in late-type dwarfs \citep{2001A&A...367..218B, 2002ApJ...577..433G, 2003MNRAS.346..473K, 2009ApJ...701.1534A} and dramatic changes in the spectral energy distributions between the L dwarf and T dwarf classes  \citep{2000ApJ...536L..35L, 2001ApJ...556..357A, 2001ApJ...556..872A, 2006ApJ...640.1063B}. 
Non-equilibrium chemistry and atmospheric circulation
\citep{2000ApJ...531..438B, 2002Icar..155..393L, 2006ApJ...647..552S} coupled with the complex processes of condensate grain formation, growth, circulation and evaporation \citep{2006A&A...455..325H, 2008ApJ...675L.105H} makes the formation, evolution and influence of condensate clouds in low-temperature atmospheres among the outstanding problems in brown dwarf astrophysics today.

The disappearance of photospheric condensates at
the transition between the L dwarf and T dwarf classes is one particularly interesting aspect of brown dwarf clouds.  This transition occurs over a narrow
range of temperatures ($\Delta${\teff} $\approx$ 200~K) and luminosities ($\Delta${\lbol} $\approx$ 0.2~dex; \citealt{2000AJ....120..447K, 2004AJ....127.3516G, 2007ApJ...659..655B}), 
and is accompanied by a temporary brightening at 1~$\micron$ (the ``J-band bump'';  \citealt{2002AJ....124.1170D, 2003AJ....126..975T,2004AJ....127.2948V}) and enhanced rates of multiplicity \citep{2006ApJS..166..585B}.  These effects appear to arise from 
the rapid depletion of condensate clouds at the L dwarf/T dwarf transition \citep{2006ApJ...647.1393L,2007ApJ...659..655B}, although the driving mechanism for that depletion is inadequately explained by current cloud models \citep{2006ApJ...640.1063B, 2008ApJ...678.1372C, 2008ApJ...689.1327S}.
Whether arising from the fragmentation of cloud structures \citep{2001ApJ...556..872A,2002ApJ...571L.151B}, a sudden increase in sedimentation efficiency \citep{2004AJ....127.3553K, 2009ApJ...702..154S} or some other process remains unclear, but cloud depletion at the L dwarf/T dwarf transition has important implications on cloud evolution in other low-temperature atmospheres, such as those of extrasolar planets \citep{2005MNRAS.364..649F}.

Coeval and cospatial multiples are important laboratories for studying this transition, eliminating dependencies on age, composition and distance.  So-called ``flux-reversal'' binaries, whose components straddle the L dwarf/T dwarf transition, have verified the 1~$\micron$ brightening as an intrinsic aspect of brown dwarf evolution \citep{2003AJ....125.3302G,2004ApJ...604L..61C,2006ApJS..166..585B,2006ApJ...647.1393L, 2008ApJ...685.1183L}.
Yet resolved spectroscopy of the components of these systems has been difficult to obtain,  due largely to the close separations typical of brown dwarf multiples ($a$ $\lesssim$ 20~AU; \citealt{2007ApJ...668..492A}).  In this article, we present a comprehensive analysis of two binaries whose components are suspected to straddle this transition: {\namea} (hereafter {\namesha}) and {\nameb} (hereafter {\nameshb}).  Both were resolved with the {\em Hubble Space Telescope} (HST)
Wide Field Planetary Camera 2 (WFPC2; \citealt{2001AJ....121..489R, 2003AJ....125.3302G}) and noted as potential L dwarf/T dwarf transition binaries based on their photometric properties and late (combined-light) systemic spectral types of L6 and L7, respectively.  
Here we combine resolved near-infrared photometry obtained with the HST/Near Infrared Camera and Multi-Object Spectrometer (NICMOS) and ground-based laser guide star adaptive optics (LGSAO) imaging with combined-light, near-infrared spectroscopy to infer component spectral types, colors and absolute magnitudes.  
Observations are described in Section~2.
In Section~3 we describe point spread function (PSF) fitting of the HST data that yield relative fluxes for each binary in five spectral bands spanning 1.0--1.8~$\micron$.   In Section~4 we describe our spectral fitting procedure and determine component classifications, colors and absolute magnitudes.
In Section~5 we compare these measures to current absolute magnitude/spectral type and absolute magnitude/color trends.  In Section~6 we discuss our results in the context of cloud evolution at the L dwarf/T dwarf transition, and explore the possibility of higher-order multiplicity in the case of {\namesha}.  
Results are summarized in Section~7.

\section{Observations}

\subsection{Targets}

The empirical properties of the two binaries examined here are summarized in Table~\ref{tab_properties}.
{\namesha} was initially identified in the Two Micron All Sky Survey (2MASS; \citealt{2006AJ....131.1163S}) by \citet{1999ApJ...519..802K}, and selected to be the prototype for the L6 spectral subclass.  Its optical spectrum exhibits 6708~{\AA} Li~I absorption, indicating component masses below $\sim$0.06~{\msun} \citep{1992ApJ...389L..83R, 1993ApJ...404L..17M}.
This source was resolved as a 0$\farcs$16 binary system by \citet{2001AJ....121..489R} with HST/WFPC2, and has been subsequently confirmed as a common proper motion pair from multi-epoch HST and LGSAO observations \citep{2008A&A...481..757B,2010ApJ...711.1087K}.  The large difference in component brightnesses in the WFPC2 F814W band ($\Delta$F814W = 1.47$\pm$0.09; \citealt{2003AJ....126.1526B}) has 
suggested a late-type L or early-type T dwarf secondary, although this component has been a persistent outlier in color magnitude diagrams.
Astrometric parallax measurements by \citet{2002AJ....124.1170D} and \citet{2004AJ....127.2948V} find discrepant distances of 25.6$\pm$2.3~pc and 38$\pm$6~pc, differing by nearly 2$\sigma$.  \citet{faherty0850} identified an unrelated background source that may have skewed the \citet{2002AJ....124.1170D} astrometric measurements, and report a preliminary parallax that is intermediate between these two values.  We adopt the \citet{2004AJ....127.2948V} for this study, as was also used by \citet{2010ApJ...711.1087K}.  \citet{faherty0850} also identified a widely-separated, common proper motion and common distance companion to {\namesha}, the M5 + M6 binary NLTT 20346AB at a projected separation of 4$\farcm$1 (7700~AU).  H$\alpha$ and X-ray emission in these M dwarfs, coupled with {\lii} absorption in {\namesha}, indicate a relatively young age of 0.25--1.5~Gyr for the combined system, and relatively low masses for the {\namesha} components: 0.04$\pm$0.02~{\msun} and 0.03$\pm$0.01~{\msun} based on evolutionary models \citep{faherty0850}.  These mass estimates are consistent with weak empirical constraints by \citet{2010ApJ...711.1087K}, M$_{total}$ = 0.2$\pm$0.2~{\msun}, as inferred from incomplete astrometric monitoring of its orbit.

\begin{deluxetable*}{lccl}
\tabletypesize{\scriptsize}
\tablecaption{Properties of {\namea} and {\nameb}\label{tab_properties}}
\tablewidth{0pt}
\tablehead{
\colhead{Parameter} &
\colhead{\namesha} &
\colhead{\nameshb} &
\colhead{Reference} \\
}
\startdata
Optical Spectral Type\tablenotemark{a} & L6 & L7 & 1,2 \\
NIR Spectral Type\tablenotemark{a,b} & L7$\pm$2 & L6$\pm$1 & 3 \\
MKO $J$ & 16.20$\pm$0.03 & 15.90$\pm$0.08\tablenotemark{c} & 3,4 \\
MKO $J-K$ & 1.85$\pm$0.04 & 2.01$\pm$0.09\tablenotemark{c} & 3,4 \\
Distance (pc) & 38$\pm$6\tablenotemark{d} & 24.1$\pm$1.9 & 5 \\
{\vtan} ({\kms}) & 26.6$\pm$4.5 & 5.1$\pm$0.9 & 5 \\
Separation (mas)\tablenotemark{e} & 157$\pm$3 & 131$\pm$3 & 6 \\
\phm{Separation} (AU) & 6.0$\pm$0.9 & 3.2$\pm$0.3 & 5,6 \\
Position Angle ($\degr$)\tablenotemark{e} & 114.7$\pm$0.3 & 27.6$\pm$1.2 & 6 \\
$\Delta{F814W}$ & 1.47$\pm$0.09 & 0.37$\pm$0.04\tablenotemark{f} & 6,7 \\
$\Delta{F1042M}$ & \nodata & -0.25$\pm$0.14 & 7 \\
$\Delta{K_p}$\tablenotemark{g} & 0.78$\pm$0.07 & 0.63$\pm$0.03 & 8 \\
Li~I? & Yes & No & 1,2 \\
Age (Gyr)\tablenotemark{h} & 0.25--1.5 & $\gtrsim$1.5 & 3,9 \\
Combined Mass ({\msun})\tablenotemark{i} & 0.2$\pm$0.2 & 0.15$^{+0.25}_{-0.04}$ & 8 \\
\enddata
\tablenotetext{a}{Classification of combined light spectra.}
\tablenotetext{b}{Based on spectral indices measured from SpeX spectroscopy; see Section 2.2 and Table~\ref{tab_indices}.}
\tablenotetext{c}{Synthesized from 2MASS photometry and spectrophotometric filter corrections computed from the spectrum shown in Figure~\ref{fig_nirspec}.}
\tablenotetext{d}{\citet{2002AJ....124.1170D} and \citet{faherty0850} report parallactic distances of 25.6$\pm$2.3~pc and 29$\pm$7~pc for {\namesha}, both closer than the \citet{2004AJ....127.2948V} measurement. The Faherty et al.\ study propose that the differences may arises from a contaminant background source skewing center-of-light measurements.  We adopt the Vrba et al.\ distance for consistency, and note that an uncertainty-weighted mean of this measurement and that of Faherty et al., 36$\pm$5 pc, is fully consistent with the adopted value.}
\tablenotetext{e}{At epoch 2000 February 1 (UT) for {\namesha} and 2000 August 12 (UT) 
from {\nameshb}.}
\tablenotetext{f}{\citet{2003AJ....126.1526B} report $\Delta$F814W = 0.66$\pm$0.11 for this system based on the same dataset, a 2.5$\sigma$ difference.}
\tablenotetext{g}{Uncertainty-weighted averages of multi-epoch flux ratio measurements from \citet{2010ApJ...711.1087K}, using both statistical and PSF-matching systematic uncertainties.}
\tablenotetext{h}{Based on presence/absence of {\lii}, and for {\namesha} activity diagnostics in its co-moving companion NLTT 20346AB \citep{faherty0850}.}
\tablenotetext{i}{Based on partial coverage of astrometric orbits.}
\tablerefs{(1) \citet{1999ApJ...519..802K}; (2) \citet{2000AJ....120..447K}; (3) This paper; (4) \citet{2006AJ....131.1163S}; (5) \citet{2004AJ....127.2948V}; (6) \citet{2003AJ....126.1526B}; (7) \citet{2003AJ....125.3302G}; (8) \citet{2010ApJ...711.1087K}; (9) \citet{faherty0850}.}
\end{deluxetable*}

{\nameshb} was also identified in 2MASS by \citet{2000AJ....120..447K} and classified L7 based on its optical spectrum.  This source shows no indication of {\lii} absorption and hence its primary is inferred to have a mass greater than 0.06~{\msun}. 
\citet{2003AJ....125.3302G} and \citet{2003AJ....126.1526B} identified this system as a 0$\farcs$13 binary based on HST/WFPC2 observations. The former study found the ``secondary'' to be fainter in the F814W band and brighter in the F1042M band, the first reported example of a L/T flux reversal binary.  Multi-epoch HST and LGSAO observations reported in \citet{2008A&A...481..757B} and \citet{2010ApJ...711.1087K} have verified the common proper motion of the components, with the latter study finding an orbital period of 31$\pm$12~yr, semimajor axis of 5.3$\pm$0.8~AU (based on the parallax distance measurement of 24.1$\pm$1.9~pc;  \citealt{2004AJ....127.2948V}) and total system mass of 0.15$^{+0.25}_{-0.04}$~{\msun}.  However, like {\namesha}, the orbit of this system has not been fully sampled.
Assuming a primary temperature of $\approx$1450~K based on the combined light optical spectral type \citep{2009ApJ...702..154S, 2010ApJ...711.1087K}, the absence of {\lii} indicates a system age of at least 1.5~Gyr, according to the evolutionary models of \citet{1997ApJ...491..856B} and \citet{2008ApJ...689.1327S}.

\subsection{IRTF/SpeX Observations}

Low resolution near-infrared
spectral data for {\namesha} and {\nameshb} were obtained with
the 3m NASA Infrared Telescope Facility (IRTF)
SpeX spectrograph \citep{2003PASP..115..362R}
on  2008 January 8 (UT) and 2006 August 21 (UT), respectively.  
Conditions on both nights were mostly clear, with light cirrus during the 2006 August observations; seeing was $\sim$0$\farcs$6 at $J$-band in both observations.
We employed the prism-dispersed mode of SpeX 
which provides 0.75--2.5~$\micron$ continuous
spectroscopy at a resolution {\ldl} $\approx 120$ for the 0$\farcs$5
slit, and dispersion of 20--30~{\AA}~pixel$^{-1}$.
The slit was aligned to the parallactic angle in all observations.
{\namesha} was observed at an airmass of 1.13, with four exposures of 150~s each obtained in an ABBA dither pattern along the slit. {\nameshb} was observed in a similar manner, at  an airmass of 1.06 and with four exposures of 180~s each.
After each target observation, 
A0 V stars HD~74721 (V = 8.71) and HD~164899 (V = 7.91)
were observed with identical instrument settings and
at a similar airmass for telluric absorption and flux calibration.  These were followed
by internal flat field and Ar arc lamps for pixel response and wavelength calibration.
Data were reduced using the SpeXtool package, version 3.3
\citep{2003PASP..115..389V,2004PASP..116..362C} using standard settings;
see \citet{2006AJ....131.1007B} for details.

\begin{figure}[ht]
\centering
\epsscale{1.0}
\plotone{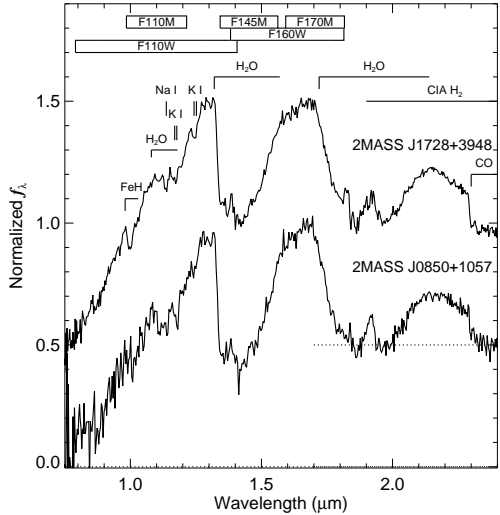}
\caption{Near-infrared spectra of {\namesha} (bottom) and {\nameshb} (top) obtained with IRTF/Spex.
Data are normalized at the 1.3~$\micron$ spectral peaks, with the spectrum of {\nameshb} vertically
offset for clarity (dotted line).  Major spectral features characteristic of L dwarf spectra
are labeled.  Also indicated are the wavelength ranges spanned by the F110M, F110W, F145M, F160W and F170M HST/NICMOS filters.
\label{fig_nirspec}}
\end{figure}

The reduced spectra for 
both sources are shown in Figure~\ref{fig_nirspec}.
The data exhibit classic signatures of L dwarf near-infrared spectra, including a
steep 0.8--1.2~$\micron$ spectral slope; FeH absorption at 1.0~$\micron$;
unresolved {\ki} and {\nai} doublets
at 1.13, 1.17 and 1.22~$\micron$; deep {\wat} absorption bands centered at 1.4
and 1.9~$\micron$; and CO absorption at 2.3~$\micron$ \citep{2001AJ....121.1710R, 2003ApJ...596..561M, 2005ApJ...623.1115C}.
Their overall near-infrared spectral energy distributions are fairly red, consistent with
their $J-K_s$ colors and indicative of cloud opacity capping the 1.25~$\micron$ and 1.65~$\micron$
spectral peaks \citep{2001ApJ...556..872A}.  There is no obvious indication of {\meth} absorption at
either 1.6~$\micron$ or 2.2~$\micron$, commonly seen
in the combined light spectra of L dwarf/T dwarf pairs
(e.g., \citealt{2004ApJ...604L..61C, 2007AJ....134.1330B}).  

\begin{deluxetable}{lcccc}
\tabletypesize{\scriptsize}
\tablecaption{Near-Infrared Spectral Index Measurements\label{tab_indices}}
\tablewidth{0pt}
\tablehead{
 & \multicolumn{2}{c}{\namesha} & \multicolumn{2}{c}{\nameshb} \\
\cline{2-3} \cline{4-5}
\colhead{Index} &
\colhead{Value} &
\colhead{SpT} &
\colhead{Value} &
\colhead{SpT} \\
}
\startdata
{\wat}-J & 0.673$\pm$0.014 & L8.0$\pm$0.4 & 0.712$\pm$0.009 & L6.9$\pm$0.3 \\
%{\meth}-J & 0.728$\pm$0.011 & \nodata & 0.752$\pm$0.008 & \nodata \\
{\wat}-H & 0.652$\pm$0.010 & L8.2$\pm$0.4 & 0.723$\pm$0.005 & L5.6$\pm$0.2 \\
%{\meth}-H & 1.107$\pm$0.012 & \nodata & 1.070$\pm$0.006 & \nodata \\
%{\wat}-K & 0.747$\pm$0.011 & L5.7$\pm$0.5 & 0.755$\pm$0.005 & L5.4$\pm$0.2 \\
{\meth}-K & 0.987$\pm$0.009 & L4.8$\pm$0.3 & 0.955$\pm$0.004 & L5.9$\pm$0.2 \\
\cline{1-5}
Mean SpT &  & L7$\pm$2 &  & L6$\pm$1 \\
\enddata
\tablecomments{Spectral indices and index-spectral type relations are defined in \citet{2006ApJ...637.1067B} and \citet{2007ApJ...659..655B}, respectively.  Mean values and uncertainties in indices and index types were calculated from 1000 realizations of each spectrum randomly varied according its noise spectrum.  Overall mean types and their uncertainties are the average and scatter of the index types, rounded off to the nearest whole subtype.}
\end{deluxetable}

We derived near-infrared spectral types for these sources using the {\wat}-J, {\wat}-H and {\meth}-K spectral indices 
defined in \citet{2006ApJ...637.1067B}, and the index-spectral type relations defined in \citet{2007ApJ...659..655B} which have a typical scatter of $\sim$1~subtype for L0--L8 dwarfs.  Values are reported in Table~\ref{tab_indices}.   The inferred near-infrared types are L7$\pm$2 for {\namesha} and L6$\pm$1 for {\nameshb}.  These types are later and earlier than the reported optical types for these sources, respectively, albeit consistent within the uncertainties.  The larger uncertainty in the near-infrared spectral type of {\namesha} is driven by a discrepant {\meth}-K index value, which indicates an L5 spectral type as compared to L8 from the {\wat} indices.  This difference may be related to the multiplicity of this system, or its young age and the corresponding low surface gravity of its components.

\subsection{HST/NICMOS Observations}

Both sources were observed in single orbits
with the HST/NICMOS NIC1 camera as part of program GO-9843 (PI Gizis).
{\namesha} was observed on 2003 November 9 (UT) and
{\nameshb} on 2003 September 7 (UT).
NIC1 is the highest-resolution camera on NICMOS with pixel scale 0$\farcs$043 and
field of view 11$\arcsec$$\times$11$\arcsec$, providing a well-sampled
PSF down to the diffraction limit at 1~$\micron$.
Both sources were observed through the wide-band filters 
F110W ($\lambda_c$ = 1.025~$\micron$, $\Delta\lambda$ = 0.6~$\micron$) and
F160W ($\lambda_c$ = 1.55~$\micron$, $\Delta\lambda$ = 0.4~$\micron$), and the medium-band filters
F110M ($\lambda_c$ = 1.1~$\micron$, $\Delta\lambda$ = 0.2~$\micron$),
F145M ($\lambda_c$ = 1.45~$\micron$, $\Delta\lambda$ = 0.2~$\micron$), and
F170M ($\lambda_c$ = 1.7~$\micron$, $\Delta\lambda$ = 0.2~$\micron$).   As shown in Figure~\ref{fig_nirspec}, the F110W and F160W filters sample the prominent $J$ and $H$ flux peaks of late-type L and T dwarf spectra, while the F110M, F145M
and F170M filters sample the 1.1~$\micron$ {\wat}, 
1.4~$\micron$ {\wat} and 1.6~$\micron$ {\meth} absorption bands, 
respectively.  These filter combinations provide discriminating colors
for late-type L and T dwarfs (e.g., \citealt{2006ApJS..166..585B,2006ApJ...639.1114R}; Figure~\ref{fig_color}).
All data were acquired in MULTIACCUM mode, with two dithered exposures (1$\farcs$5 step) 
obtained in each filter.  Total integration times for each source were 256~s in F110W and F110M ({\nameshb} was observed for 288~s in F110M), 320~s in F160W, 704~s in F170M and 832~s in F145M.  Integrations were longest with the F145M filter because of the deep {\wat} absorption present at these wavelengths.

Raw images were reduced by standard pipeline processing 
(CALNICA, \citealt{1997hstc.work..223B}) using
calibration images (inflight and model reference files, circa 2002)
and photometric keywords (post-cyrogenic) current as of February 2008.
CALNICA reduction includes analog-to-digital correction, 
subtraction of bias and dark current frames, linearity correction,
correction for readout artifacts (the ``bars'' anomaly), division by an
appropriate flat field image, photometric calibration, cosmic
ray identification, and combination of MULTIACCUM frames
into a single calibrated image.  No correction was made
for the temperature dependence of the dark and flat-field reference files.  These basic calibrated images (two images
per filter per object) were used for the analysis.

\begin{figure*}[ht]
\centering
\epsscale{0.55}
\plottwo{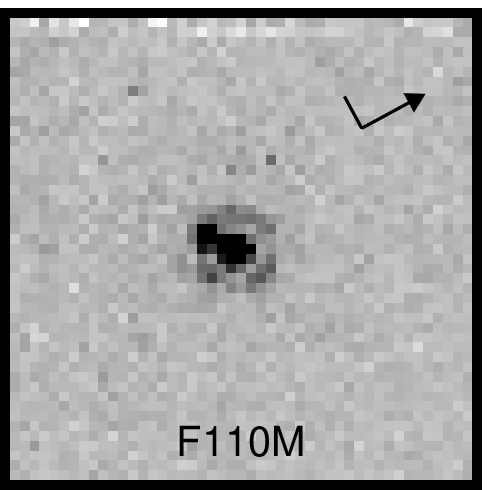}{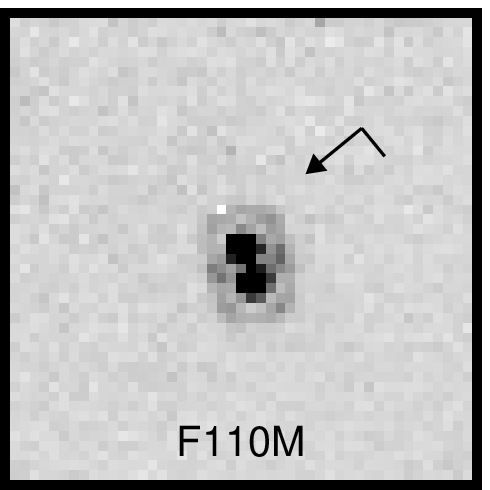} \\
\plottwo{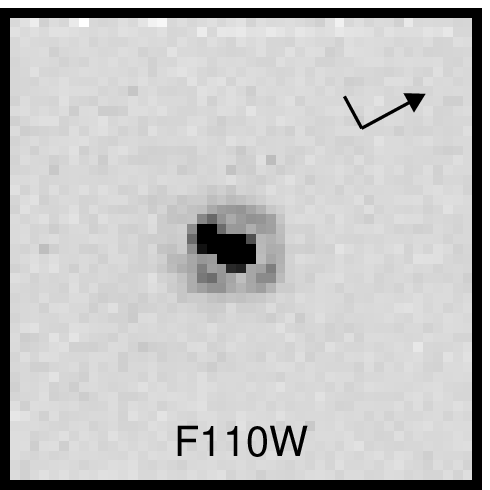}{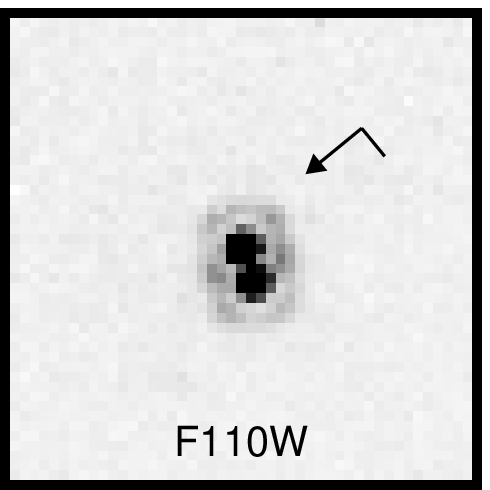} \\
\plottwo{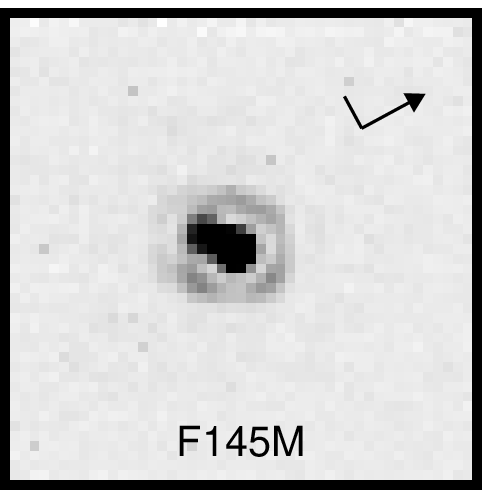}{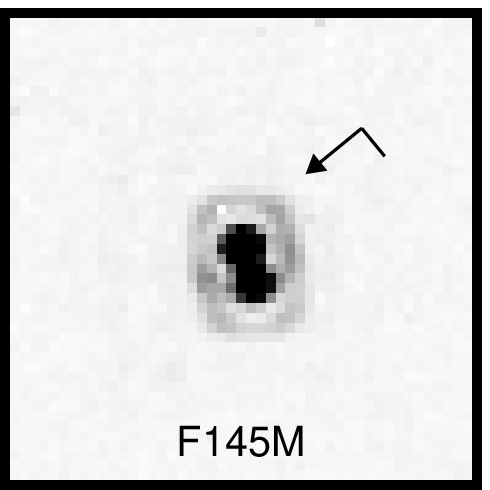} \\
\plottwo{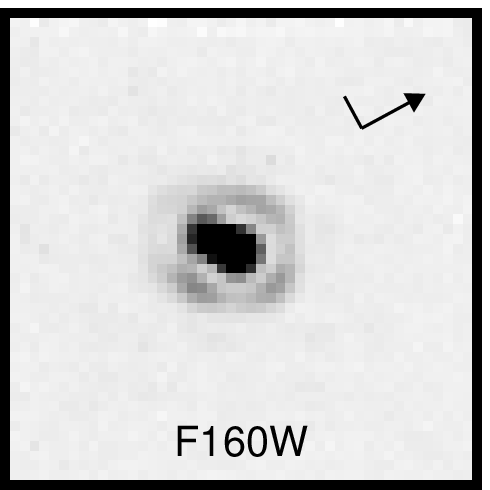}{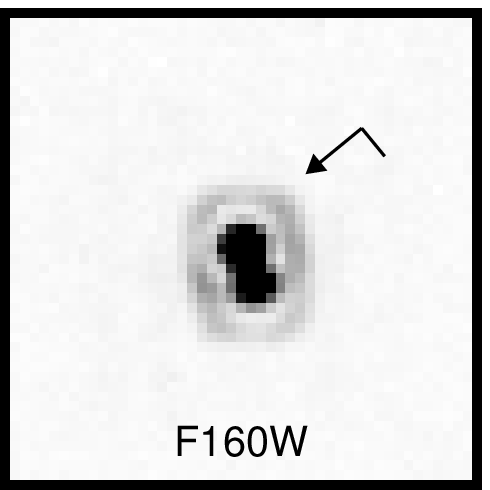} \\
\plottwo{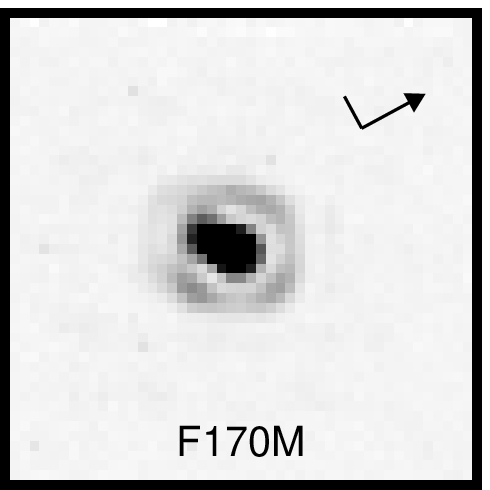}{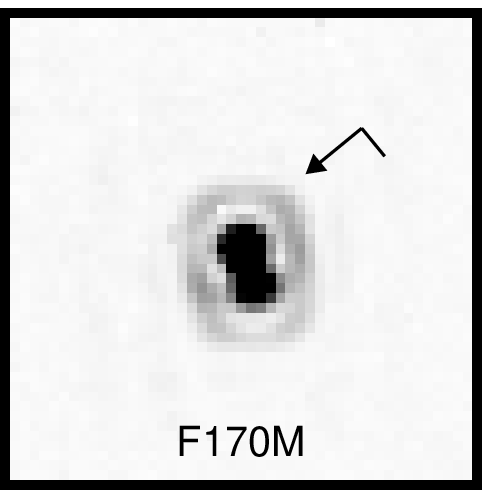}
\caption{HST/NICMOS images of {\namesha} (left) and {\nameshb} (right) in the 
F110M, F110W, F145M, F160W and F170M filters (top to bottom).  Each image is 
2$\farcs$15 (50 NIC1 pixels) on a side with orientations on the sky
indicated by the arrows (head points north, stem points east).
\label{fig_images}}
\end{figure*}

Figure~\ref{fig_images} displays subsections
of the calibrated imaging data for each binary in the five filter bands.  Two components
are resolved in each image, although the increase in PSF size 
in the longer wavelength data results in better PSF separation in the F110M and F110W images as compared to the F170M images.  There are notable differences in the relative component fluxes between these two sources.  The eastern component of {\namesha} is significantly fainter than the western component in all five filter bands, while the two components of {\nameshb} are roughly equal in brightness.  We measured aperture photometry for the combined light of each source using the IRAF\footnote{Image Reduction and Analysis Facility (IRAF;  \citealt{1986SPIE..627..733T}) is 
distributed by the National Optical
Astronomy Observatories, which are operated by the Association of
Universities for Research in Astronomy, Inc., under cooperative
agreement with the National Science Foundation.} PHOT routine, employing a 15 pixel 
radius aperture centered on the brightest component and a 20-25 pixel
annulus to measure the median background.  Count rates were converted to Vega magnitudes on the CIT system using photometric conversion parameters listed in
the NICMOS Data Handbook version 8.0\footnote{\url{http://www.stsci.edu/hst/nicmos/documents/handbooks/DataHandbookv8}.},
but without applying aperture corrections.  Values are listed in Table~\ref{tab_nicmos}, where the uncertainties include shot noise and background uncertainties, as well as a standard 5\% calibration uncertainty.
Relative photometric measurements are discussed below.

\begin{deluxetable*}{lcccc|cccc}
\tabletypesize{\scriptsize}
\tablecaption{HST/NICMOS Photometry and Astrometry\label{tab_nicmos}}
\tablewidth{0pt}
\tablehead{
 &
\multicolumn{4}{c|}{\namesha} &
\multicolumn{4}{c}{\nameshb} \\
\colhead{Parameter} & \colhead{AB} & \colhead{$\Delta$} & \colhead{A} & \colhead{B} &
\colhead{AB} & \colhead{$\Delta$} & \colhead{A} & \colhead{B} \\
}
\startdata
Epoch &  \multicolumn{4}{c|}{2003 Nov 9}  & \multicolumn{4}{c}{2003 Sep 7}  \\
F110W  & 17.38$\pm$0.07 & 1.15$\pm$0.03 &  17.70$\pm$0.07  & 18.99$\pm$0.08 &  16.76$\pm$0.05 & 0.322$\pm$0.015  & 17.36$\pm$0.05 & 17.68$\pm$0.05 \\
F110M  & 17.49$\pm$0.06 & 1.176$\pm$0.015 & 17.81$\pm$0.06 & 18.99$\pm$0.08 &  16.81$\pm$0.07 & 0.25$\pm$0.03 & 17.44$\pm$0.07 & 17.69$\pm$0.07 \\
F145M  & 16.48$\pm$0.05 & 1.000$\pm$0.015 &  16.84$\pm$0.05 & 17.84$\pm$0.05  &  15.77$\pm$0.05 & 0.517$\pm$0.014 & 16.29$\pm$0.05 & 16.81$\pm$0.05 \\
F160W  & 15.70$\pm$0.05 & 0.90$\pm$0.05 &  16.09$\pm$0.05 & 16.99$\pm$0.06 &  15.08$\pm$0.05 & 0.461$\pm$0.016 & 15.63$\pm$0.05 & 16.09$\pm$0.05 \\
F170M  & 15.36$\pm$0.05 & 0.89$\pm$0.07 &  15.76$\pm$0.05 & 16.65$\pm$0.07  &  14.77$\pm$0.05 & 0.462$\pm$0.014 & 15.32$\pm$0.05 & 15.78$\pm$0.05 \\
F110W-F160W  & 1.68$\pm$0.09 & \nodata & 1.61$\pm$0.09 & 1.86$\pm$0.10 & 1.68$\pm$0.07 & \nodata & 1.74$\pm$0.07 & 1.60$\pm$0.07 \\
F110W-F170M  & 2.02$\pm$0.09 & \nodata & 1.95$\pm$0.09 & 2.21$\pm$0.10 & 1.99$\pm$0.07 & \nodata & 2.05$\pm$0.07 & 1.91$\pm$0.07 \\
$\rho$ (mas) & \multicolumn{4}{c|}{132$\pm$5} & \multicolumn{4}{c}{158$\pm$5} \\
\phm{$\rho$} (AU) &  \multicolumn{4}{c|}{5.0$\pm$0.8} & \multicolumn{4}{c}{3.8$\pm$0.3} \\
PA ($\degr$) & \multicolumn{4}{c|}{126.1$\pm$0.7} & \multicolumn{4}{c}{68.0$\pm$0.5} \\
\enddata
\tablecomments{Photometry in Vega magnitudes on the CIT system based on conversions given in
the NICMOS Data Handbook version 8.0.}
\end{deluxetable*}

\section{HST/NICMOS PSF Fitting Analysis}

\subsection{Method}

Relative photometry for the components of each binary were determined using a PSF-fitting algorithm similar to that described in \citet{2006ApJS..166..585B}.
Models for each binary image were generated using PSFs calculated with the {\em Tiny Tim} package, version 6.3 \citep{1995ASPC...77..349K}. PSFs were generated for each filter passband (assuming post-cryocooler aberrations for NIC1), and at two pointing positions to account for mirror zonal errors.  The SpeX spectra of each source were used as templates for calculating filter passband effects.  Individual PSF models were generated for a subimage size of 3$\arcsec$ $\times$ 3$\arcsec$, which was resampled at ten times the original resolution to allow for subpixel offsets.

A binary PSF model for each NICMOS image was determined using an iterative image fitting algorithm that finds optimal positions and relative fluxes. Initial guesses were found using a simple peak detection algorithm and single-PSF subtraction on a 3$\arcsec$ $\times$ 3$\arcsec$ subframe of the data image centered on the binary.
We then varied the primary position, secondary position, primary flux and secondary flux, in that order, to minimize the statistic
\begin{equation}
S^2 = \sum_i \sum_j W_{ij}(D_{ij} - M_{ij} - \langle D-M \rangle_{ij})^2
\end{equation} 
where $D$ is the data subframe, $M$ is the model image, $W$ is a masking image used to exclude bad pixels ($W$ = 0 for bad pixels), $\langle D-M \rangle$ is the mean difference (to account for residual background flux in the data) and the sum is performed over all pixels.   The iteration step in position was set at 0.1 pixels (4.3~mas) in accord with the subsampling of the PSF model; the iteration in flux values was 1\%.  Iterations were performed in a hierarchical recursive loop, and ceased when the fractional decrease in $S^2$ was less than 10$^{-8}$ for all parameters.  For each source, we adopted as our final relative flux values the mean of the two pointing frames in each filter band, and uncertainties include the difference in individual measurements and a 1\% flux sampling uncertainty.  Final separations and position angles\footnote{The position angle is measured east of north, assuming a vector that points from primary to secondary. The HST roll angles at the time of observation are included in the values, assumed to be accurate to within 0$\fdg$003 \citep{2008A&A...481..757B}.} were adopted as the mean of all frames, with uncertainties incorporating measurement scatter and a 0.1~pixel sampling uncertainty.

The close separation and overlapping PSFs of the components of these
two systems initially raised concerns that fitting biases 
for the longer-wavelength images (e.g., F170M) could lead to skewed component photometry and colors.  This was of particular concern for
{\namesha}, whose components have a larger difference in brightness.
To test our fitting program, we performed an identical analysis with 
simulated binary images generated from the model PSFs.  The simulated images
were constructed to have the same component positions and relative magnitudes as determined from the PSF fitting (see below) as well as relative magnitudes 
inferred from unconstrained spectral fits (see Section~4.1).  
In both cases, the fitting algorithm 
reproduced the input relative magnitudes in all filter bands
to within 0.01~mag, even for the F170M images of {\namesha}.

\subsection{Results}

Table~\ref{tab_nicmos} summarizes the results of the PSF fitting analysis and all of the NICMOS photometry.  
We find separations and position angles that are consistent to within 1.5$\sigma$ of those measured by \citet{2008A&A...481..757B} using the same NICMOS data.
The relative magnitudes are all positive, indicating that the primary components are brighter than the secondary components at these wavelengths for both sources.
Relative component brightnesses show distinct trends.  For {\namesha}, relative magnitudes decrease toward longer wavelengths, 
from $\Delta$F110M = 1.176$\pm$0.015 to $\Delta$F170M = 0.89$\pm$0.07, 
indicating a secondary that is redder than the primary.   This is confirmed by the 
redder broadband color of the secondary, F110W-F160W = 1.61$\pm$0.09 and 1.86$\pm$0.10 for {\namesha}A and {\namesha}B, respectively. 
In contrast, the relative magnitudes of {\nameshb} generally increase toward longer wavelengths (with the exception of the F145M band)
indicating a bluer secondary.  
This trend is also seen in resolved $JHK_p$ photometry for {\nameshb} by \citet{2010ApJ...711.1087K}.

\begin{figure*}
\centering
\epsscale{1.0}
\includegraphics[width=0.4\textwidth]{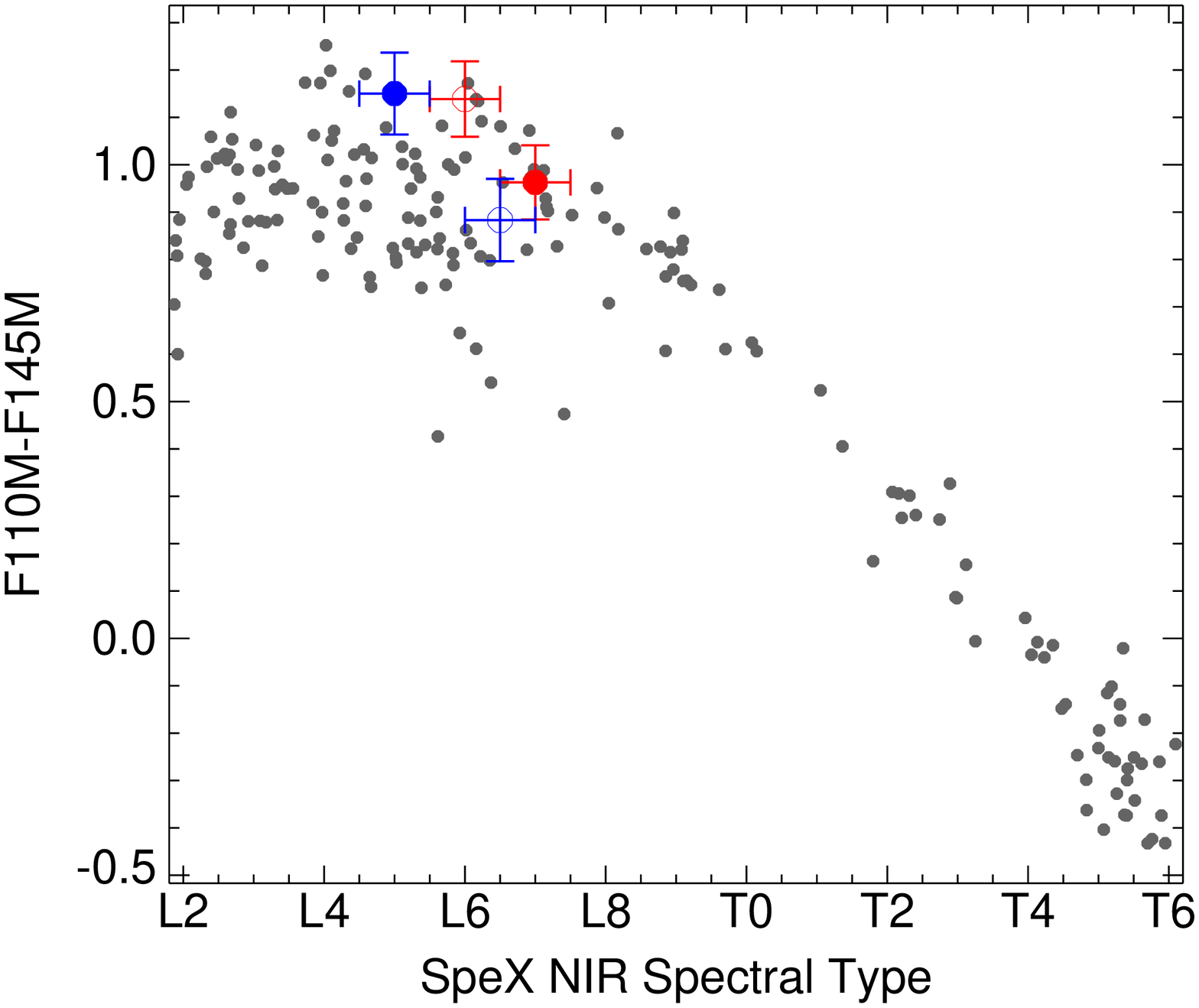}
\includegraphics[width=0.4\textwidth]{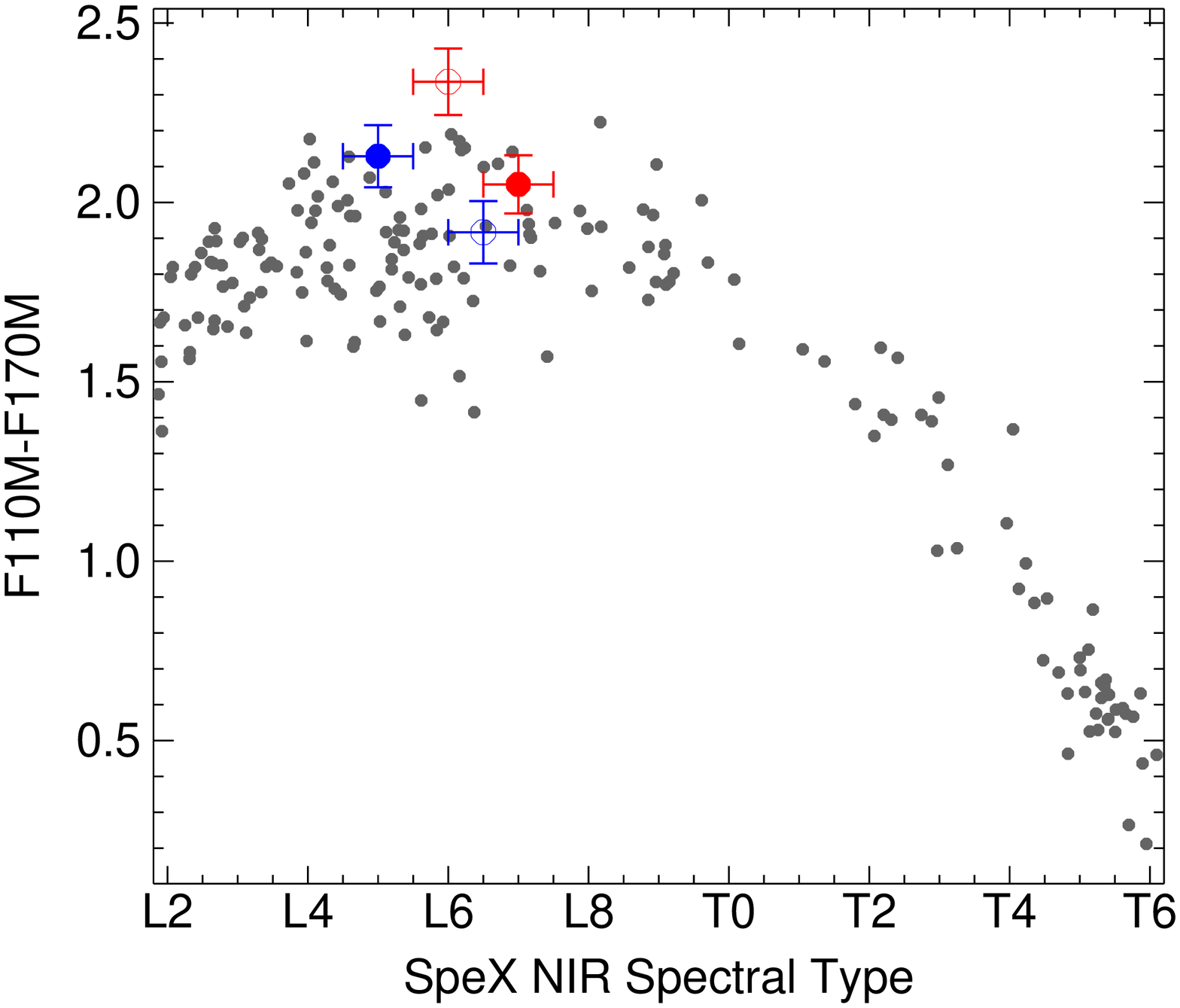}
\includegraphics[width=0.4\textwidth]{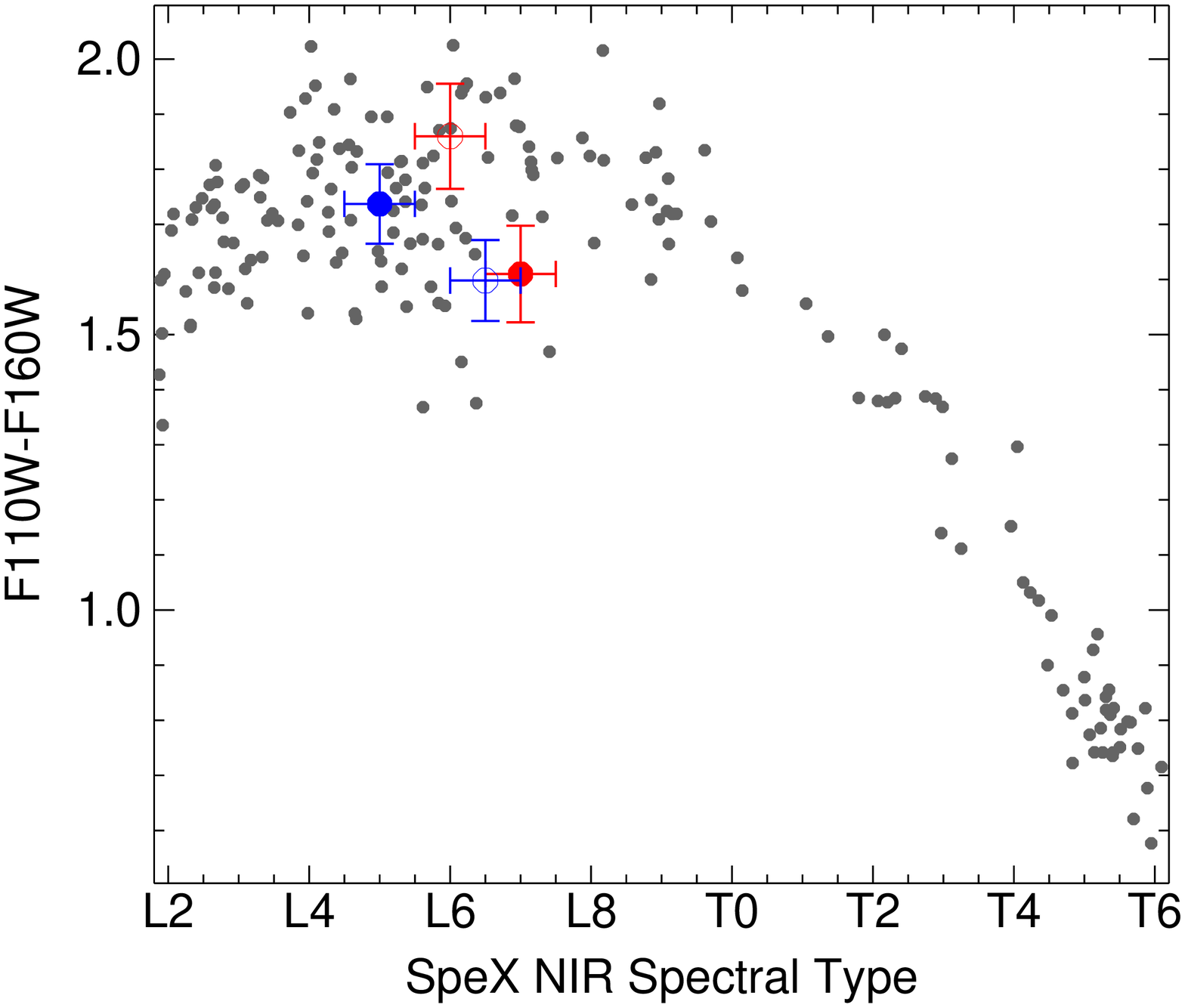}
\includegraphics[width=0.4\textwidth]{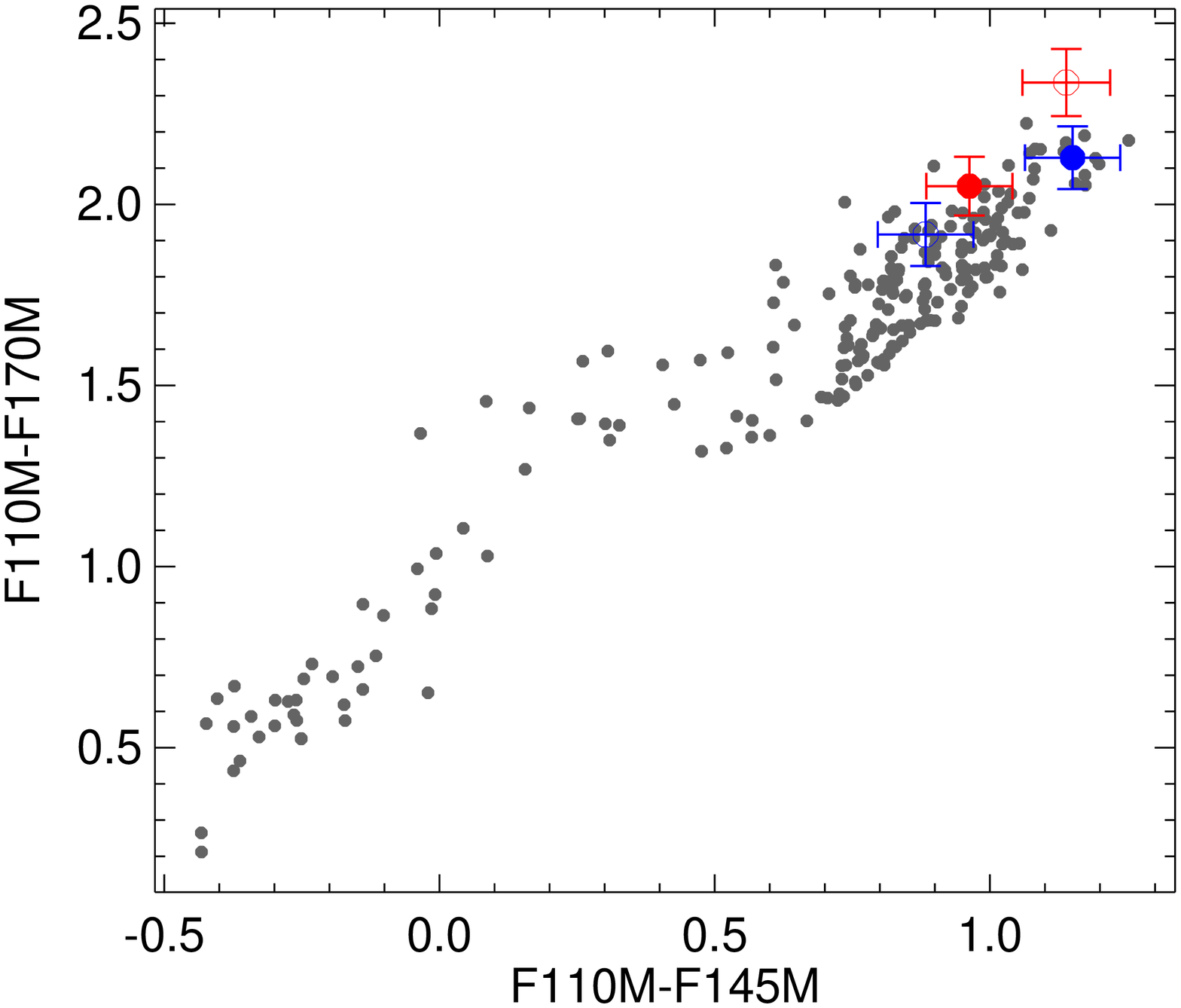}
\includegraphics[width=0.4\textwidth]{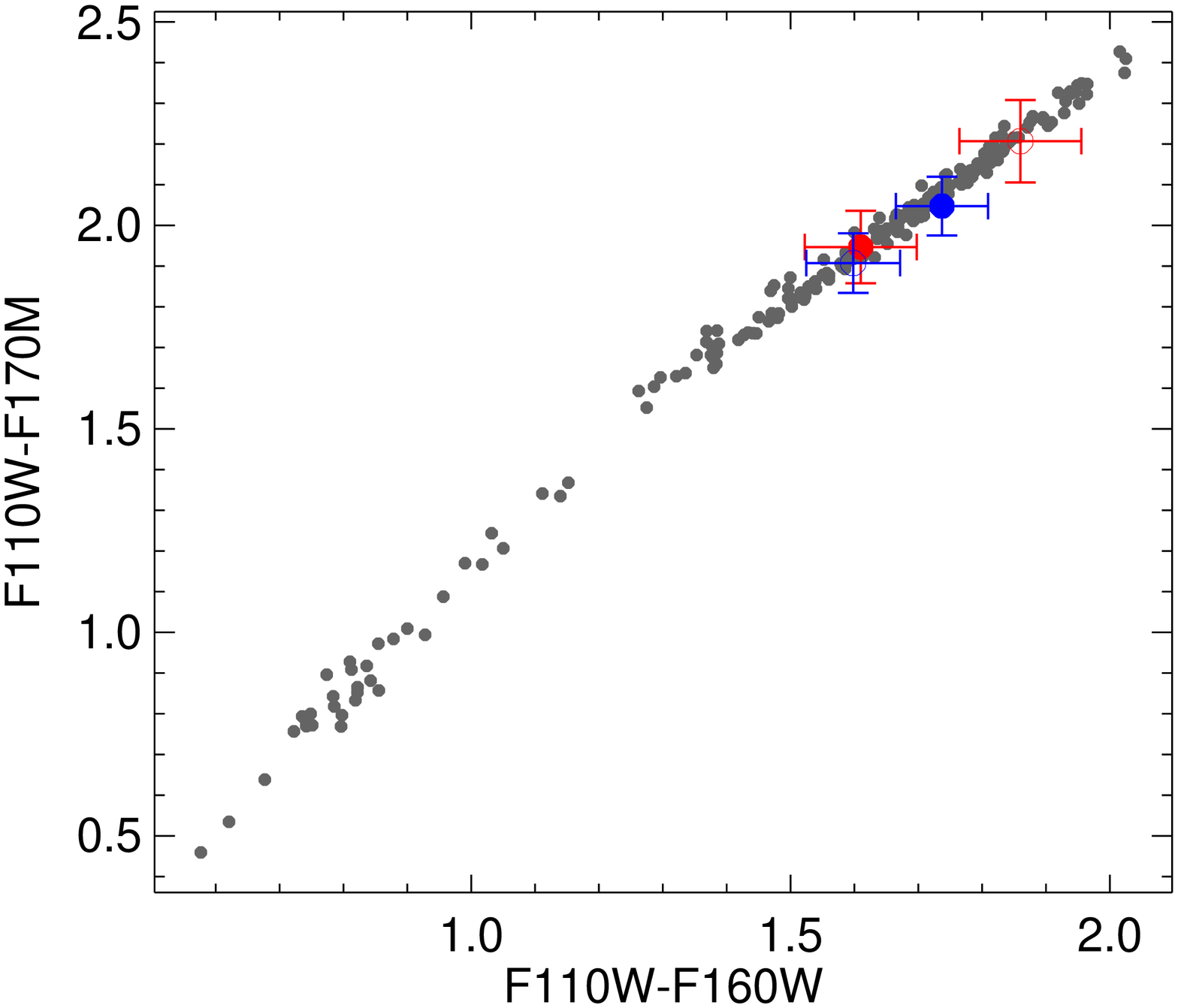}
\caption{Segregation of L and T dwarfs with NICMOS photometry, based on spectrophotometric colors computed from SpeX templates.  The first three panels show trends in narrow-band F110M-F145M and F110M-F170M colors and broadband F110W-F160W color with near-infrared (SpeX-based) spectral type.
The last two panels compare color pairs F110M-F145M versus F110M-F170M and F110W-F170M versus F110W-F160W.  
Measured colors for the individual components of {\namesha} and {\nameshb} are indicated in red and blue, respectively, with the primaries (secondaries) indicated by solid (open) circles.  In the first three panels, the components are shown with at their inferred near-infrared spectral types based on the spectral fitting analysis described in Section~4.
\label{fig_color}}
\end{figure*}

Figure~\ref{fig_color} compares the colors of the {\namesha} and {\nameshb} components to spectrophotometric estimates calculated from L2--T6 SpeX spectral templates (Section~4.1).  In general, the components span the full range of colors seen in the L dwarf templates, although {\namesha}B and (to lesser degree) {\nameshb}A are notably redder in both broadband and narrow-band colors.  F110M-F145M and F110M-F170M colors, sampling 1.4~$\micron$ {\wat} and 1.6~$\micron$ {\meth} features, indicate that none of the components have spectral types T0 or later; i.e., they are all L dwarfs.  NICMOS colors alone are unable to distinguish L subtypes for the individual components.  They are, however,  cleanly segregated in F110W-F170M and F110W-F160W colors, which are correlated for the L and T dwarf templates.  The binary components follow a sequence of {\nameshb}B, {\namesha}A, {\nameshb}A and {\namesha}B in this color-color plot, in order of increasing F110W-F170M and F110W-F160W colors. 

\section{Spectral Template Fitting Analysis}

\subsection{Method}

To infer the component near-infrared spectral types of {\namesha} and {\nameshb}, we compared their combined-light SpeX spectra to a suite of empirical
binary templates scaled to the observed relative photometry.  We followed a procedure
similar to that described in \citet{2006ApJS..166..585B}, 
drawing from a uniform sample of low-resolution, high signal-to-noise SpeX prism spectra of late-type M, L and T dwarfs.\footnote{These data were compiled from \citet{2004AJ....127.2856B,2006ApJ...639.1095B, 2006ApJ...637.1067B,2007ApJ...658..557B,2008ApJ...681..579B,2008ApJ...674..451B, 2010ApJ...710.1142B, 2004ApJ...604L..61C,2006AJ....131.1007B,2006AJ....131.2722C,2006AJ....132.2074M,2006ApJ...639.1114R,2007ApJ...659..655B,2007AJ....134.1330B,2007ApJ...658..617B,2007ApJ...655..522L,2007AJ....134.1162L,2008ApJ...686..528L,2007ApJ...654..570L,2007AJ....133.2320S,2009AJ....137..304S} and \citet{2010AJ....139.1045S}, and are available at the SpeX Prism Spectral Libraries website, \url{http://www.browndwarfs.org/spexprism}.} 
The template sample was restricted to dwarfs later than M7 based on published optical classifications for M and L dwarfs (tied largely to the schemes of \citealt{1991ApJS...77..417K,1999ApJ...519..802K}) and near-infrared classifications for T dwarfs (tied largely to the scheme of \citealt{2006ApJ...637.1067B}).  About $\sim$15\% of the M and L dwarfs in our sample have only near-infrared types in the literature, based on various schemes (e.g., 
\citealt{2001AJ....121.1710R, 2002ApJ...564..466G}).
We also computed alternate near-infrared spectral classifications directly from the SpeX data using the \citet{2001AJ....121.1710R} and \citet{2007ApJ...659..655B} index-spectral type relations, following the iterative procedure detailed in \citet{2010ApJ...710.1142B}.   
Known (resolved) binaries and sources
noted to have peculiar spectra or highly uncertain classifications ($\sigma \geq \pm$2 subtypes) were purged from the template sample.   The resulting 462 spectra of 438 sources were 
interpolated onto a common wavelength scale, and synthetic Vega magnitudes in the five NICMOS filters and MKO\footnote{Mauna Kea Observatory filter system; see \citet{2002PASP..114..180T} and \citet{2002PASP..114..169S}.} $JHKK_p$ filters were computed by convolving each spectrum and a Kurucz model of Vega with the appropriate filter transmission functions (see \citealt{2005ApJ...623.1115C}).

From these individual templates, we produced separate libraries of binary templates for {\namesha} and {\nameshb} by adding together appropriately scaled pairs.  The pairs were initially selected to have secondary component types that were no more than 2 subtypes earlier than the primary type.  This produced 137,212 unique combinations.  We then computed an uncertainty-weighted mean relative flux scaling between the two components of each system using all five NICMOS measurements and $K_p$ photometry\footnote{\citet{2010ApJ...711.1087K} also report resolved $J$- and $H$-band photometry for {\nameshb}, $\Delta{J}$ = 0.32$\pm$0.02 and $\Delta{H}$ = 0.45$\pm$0.02.  These measurements are redundant---and in agreement---with broad-band F110W and F160W NICMOS photometry, and were therefore not used in the analysis.} from \citet{2010ApJ...711.1087K}; see Table~\ref{tab_properties}.
After scaling the spectra, we required that relative synthetic magnitudes in all six filter bands agree with measured values to within 3$\sigma$.  This substantially reduced the number of ``allowable'' templates to 19,042 for {\namesha} and 1389 for {\nameshb}.  

Comparisons were then made between the source spectral data and allowed binary templates using the weighted chi-squared
statistic defined in \citet{2008ApJ...678.1372C},
\begin{equation}
G_k \equiv \sum_{\{ \lambda\} }w[{\lambda}]\left[ \frac{D[{\lambda}]-{\alpha}T[{\lambda}]}{\sigma_D[{\lambda}]} \right]^2
\label{equ_chisq}
\end{equation}
Here, $D[\lambda]$ and $T[{\lambda}]$ are the data and template spectra,
respectively; 
$\sigma_{D}[{\lambda}]$ is the uncertainty spectrum of the data;
$w[{\lambda}]$ is a vector of weights satisfying $\sum_{\{ \lambda\} }{w[{\lambda}]} = 1$; 
$\alpha$ is a scaling factor that minimizes $G_k$ (see Equation~2 in \citealt{2008ApJ...678.1372C});
and the sum is performed over the wavelength ranges $\{\lambda\}$ =
0.95--1.35~$\micron$, 1.45--1.8~$\micron$ and 2.0--2.35~$\micron$ to avoid 
regions of strong telluric absorption.  
We adopted the same weighting scheme as that used in \citet{2008ApJ...678.1372C} and \citet{2010ApJ...710.1142B}, with each
pixel weighted by its spectral width (i.e., $w_i \equiv \Delta\lambda_i$).

Templates which provided minimum values of $G_k$ were deemed to be the best fits.
However, it was generally the case that other templates gave $G_k$ values only slightly larger than this best-fit template, and are therefore statistically equivalent.
We therefore determined average component parameters (spectral types and relative magnitudes) and their uncertainties following \citet[Equations 4--6]{2010ApJ...710.1142B}, with an effective degrees of freedom $\nu$ = 253 used for weighting the parameters with the F-test statistic.

\subsection{Results}

\begin{figure*}
\centering
\epsscale{1.0}
\includegraphics[width=0.45\textwidth]{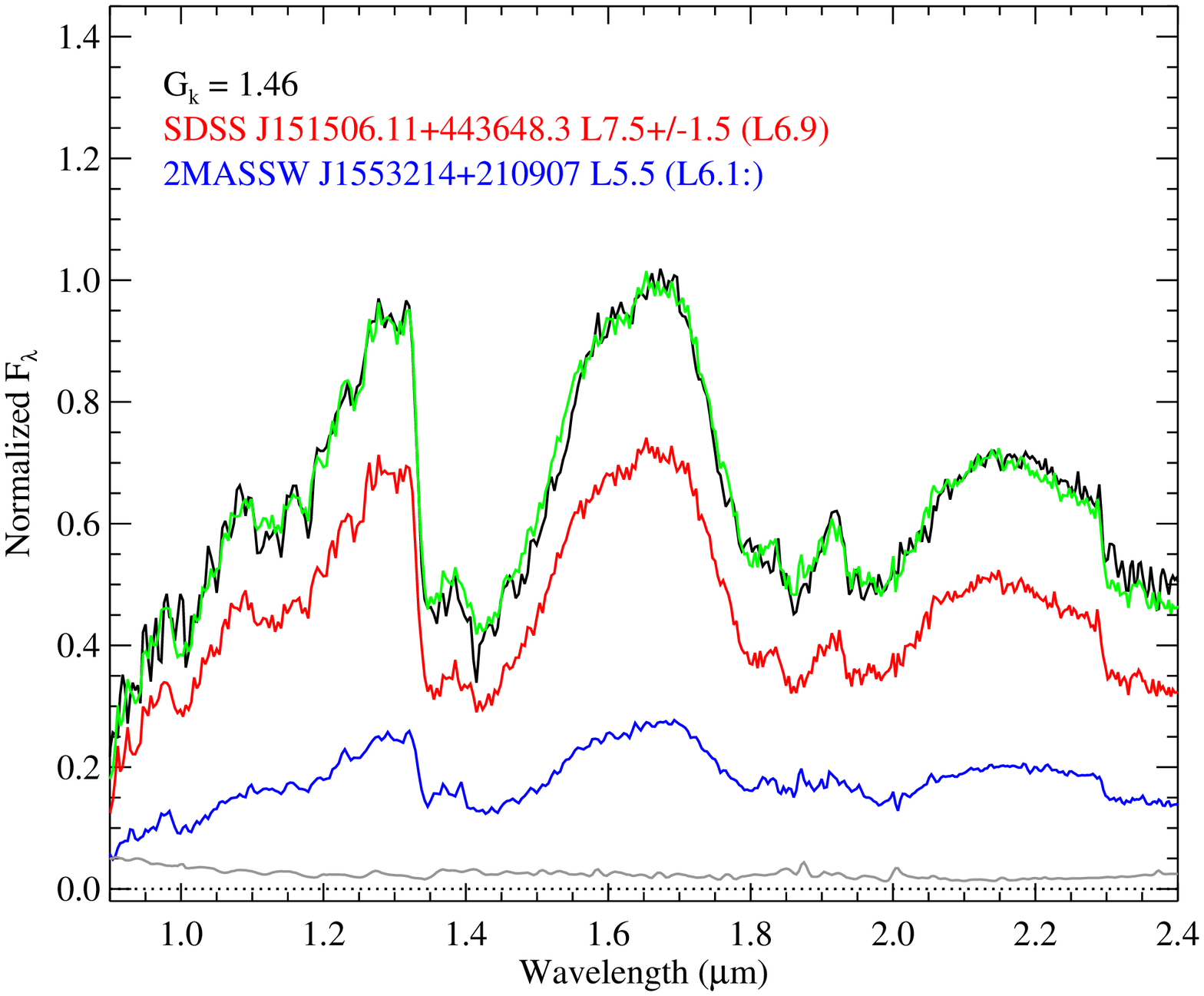}
\includegraphics[width=0.45\textwidth]{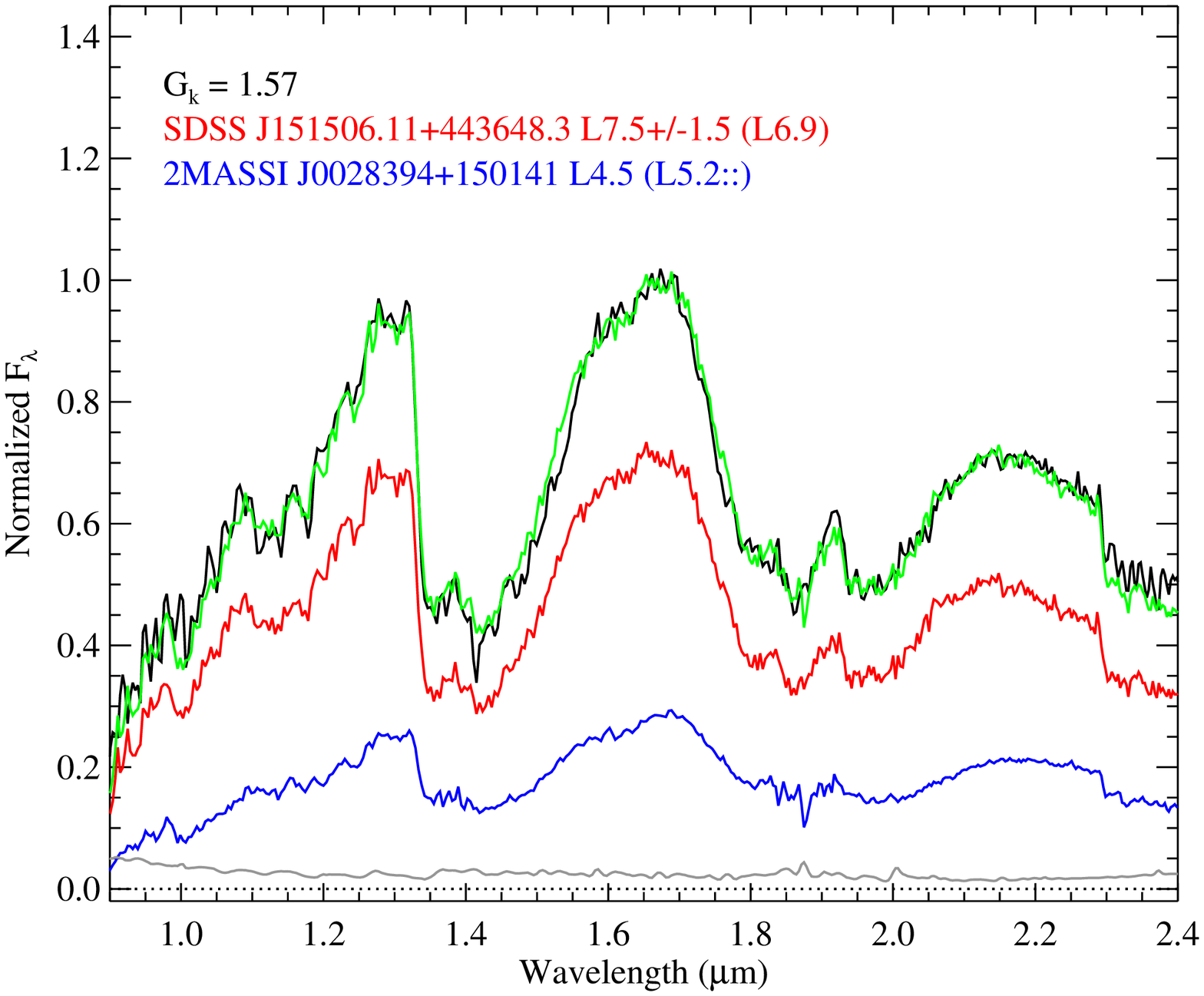}
\includegraphics[width=0.45\textwidth]{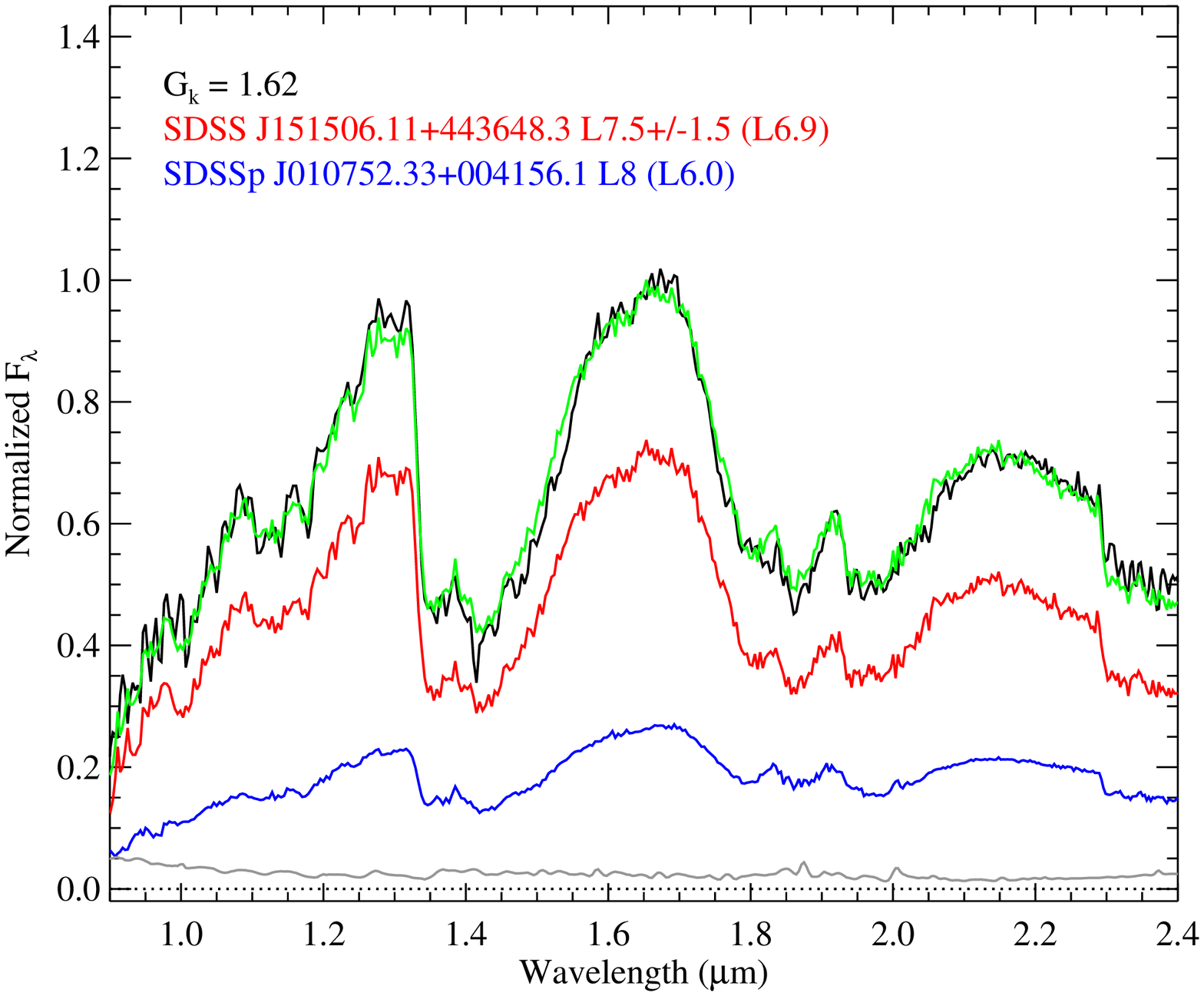}
\includegraphics[width=0.45\textwidth]{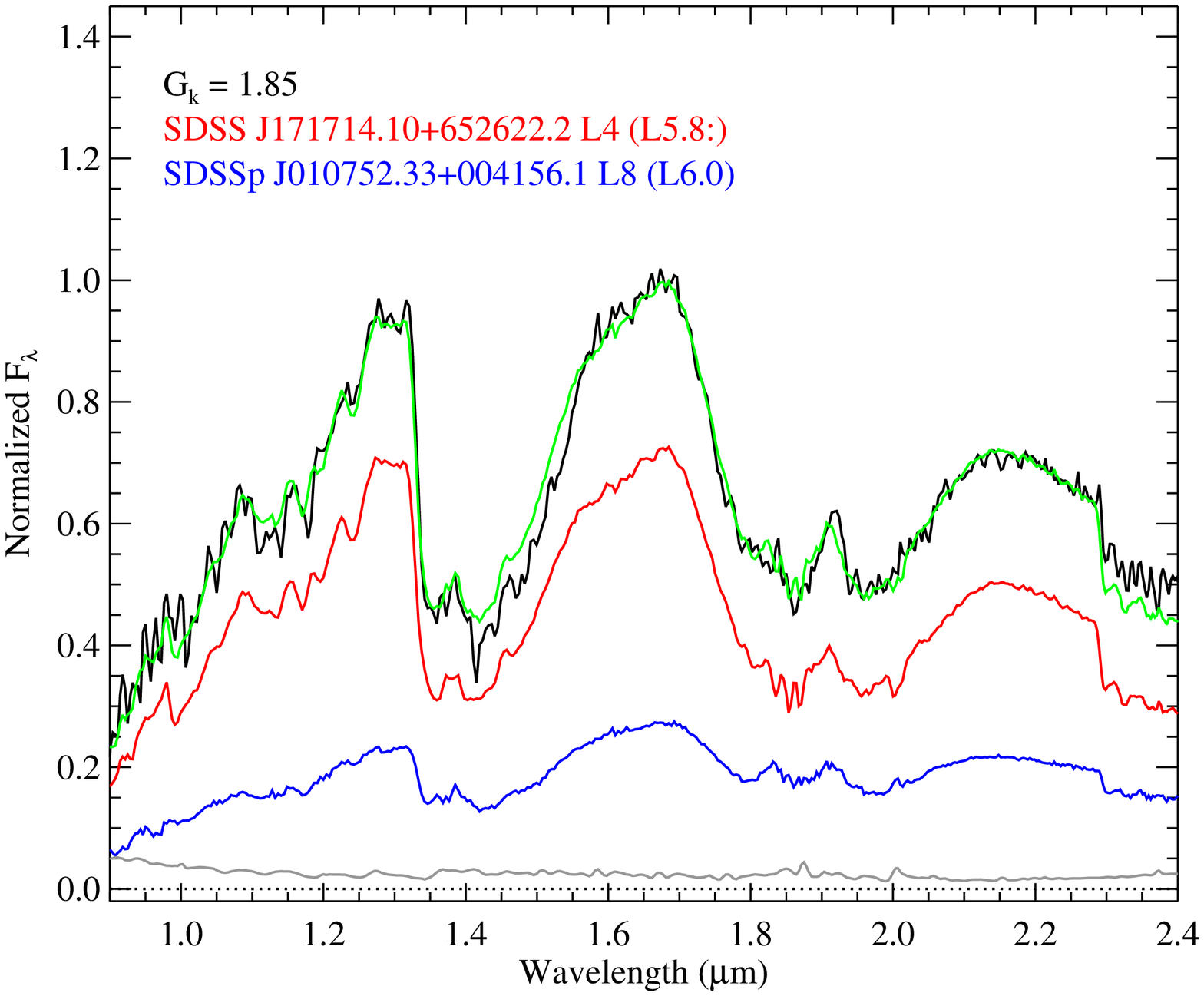}
\caption{The four best template fits for {\namesha} as constrained by HST/NICMOS and LGSAO $K_p$ photometry.
In each panel, the source spectrum is indicated by the black line, the source noise spectrum by the grey line, the best-fitting template by the green line, the best-fitting primary by the red line and the best-fitting secondary by the blue line.  Template spectra (and appropriately scaled primaries and secondaries) are
shown at their minimum $G_k$ scalings relative to the source spectrum, which is normalized at its peak spectral flux.  Template source names, literature classifications, SpeX-based classifications (in parentheses) and $G_k$ fit values are indicated.  Note that the L7.5$\pm$1.5 literature spectral type for SDSS~J151506.11+443648.3 is based on near-infrared data; all other literature types are based on optical data.
\label{fig_fit0850}}
\end{figure*}

\begin{figure*}
\centering
\epsscale{1.0}
\includegraphics[width=0.45\textwidth]{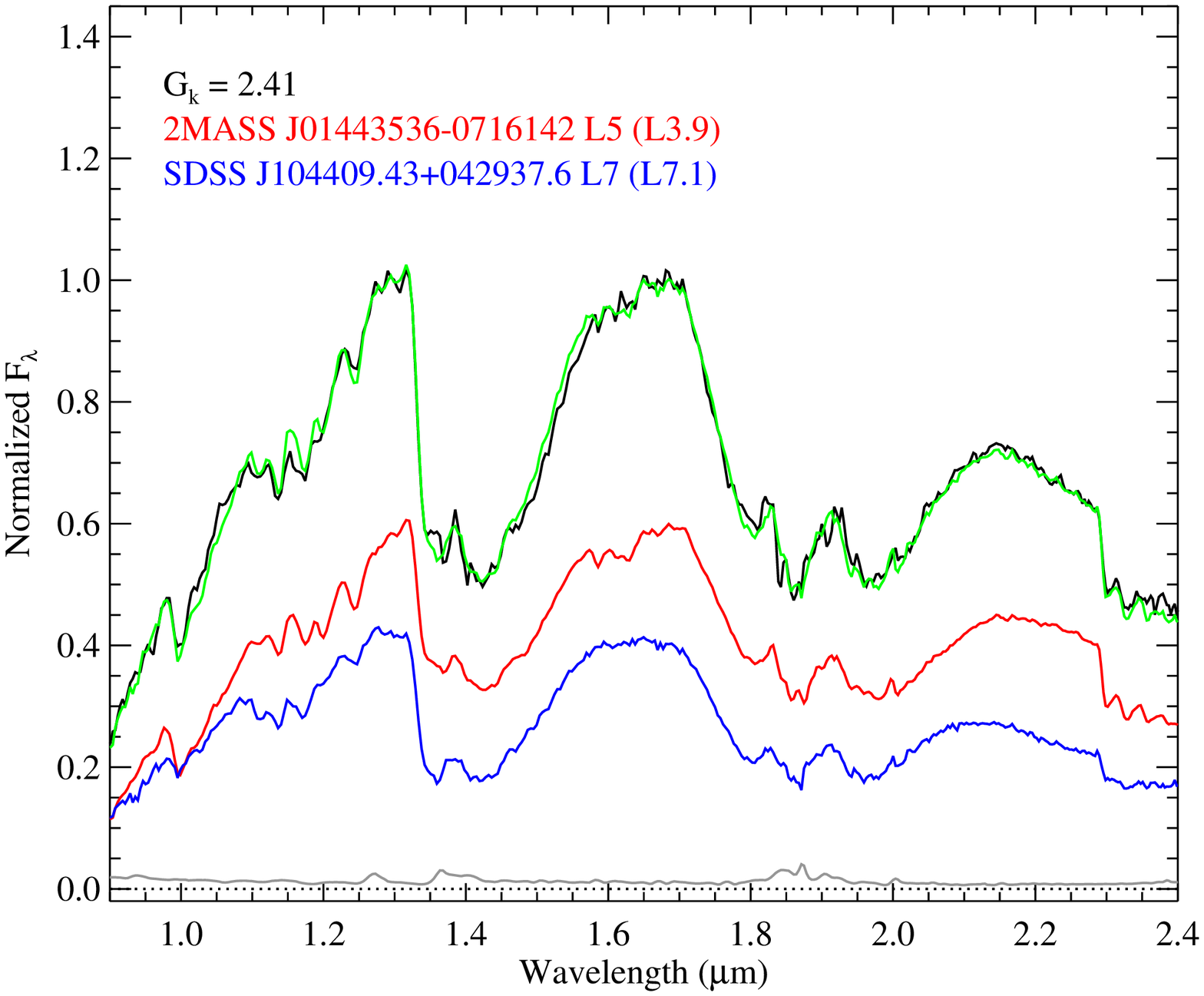}
\includegraphics[width=0.45\textwidth]{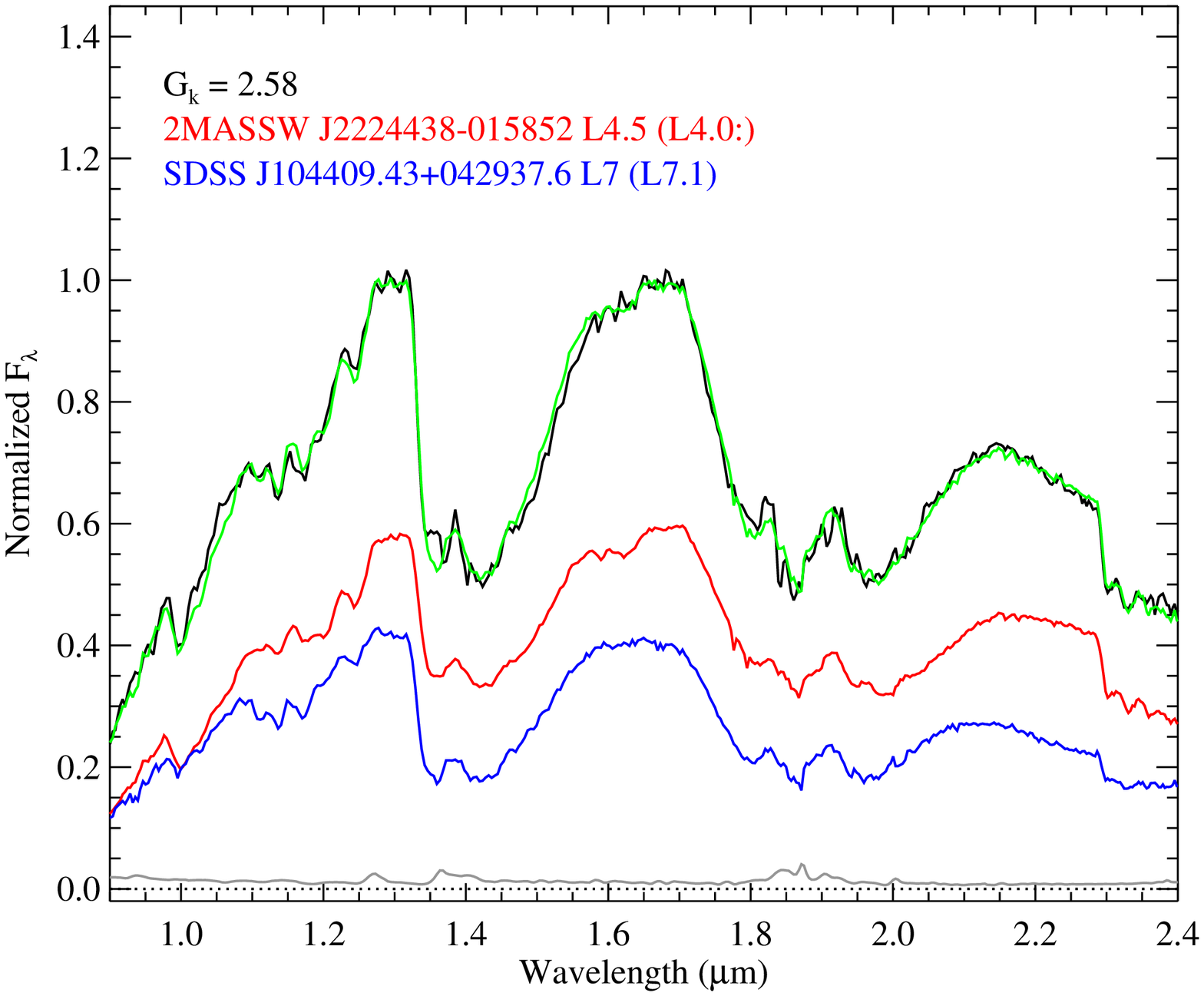}
\includegraphics[width=0.45\textwidth]{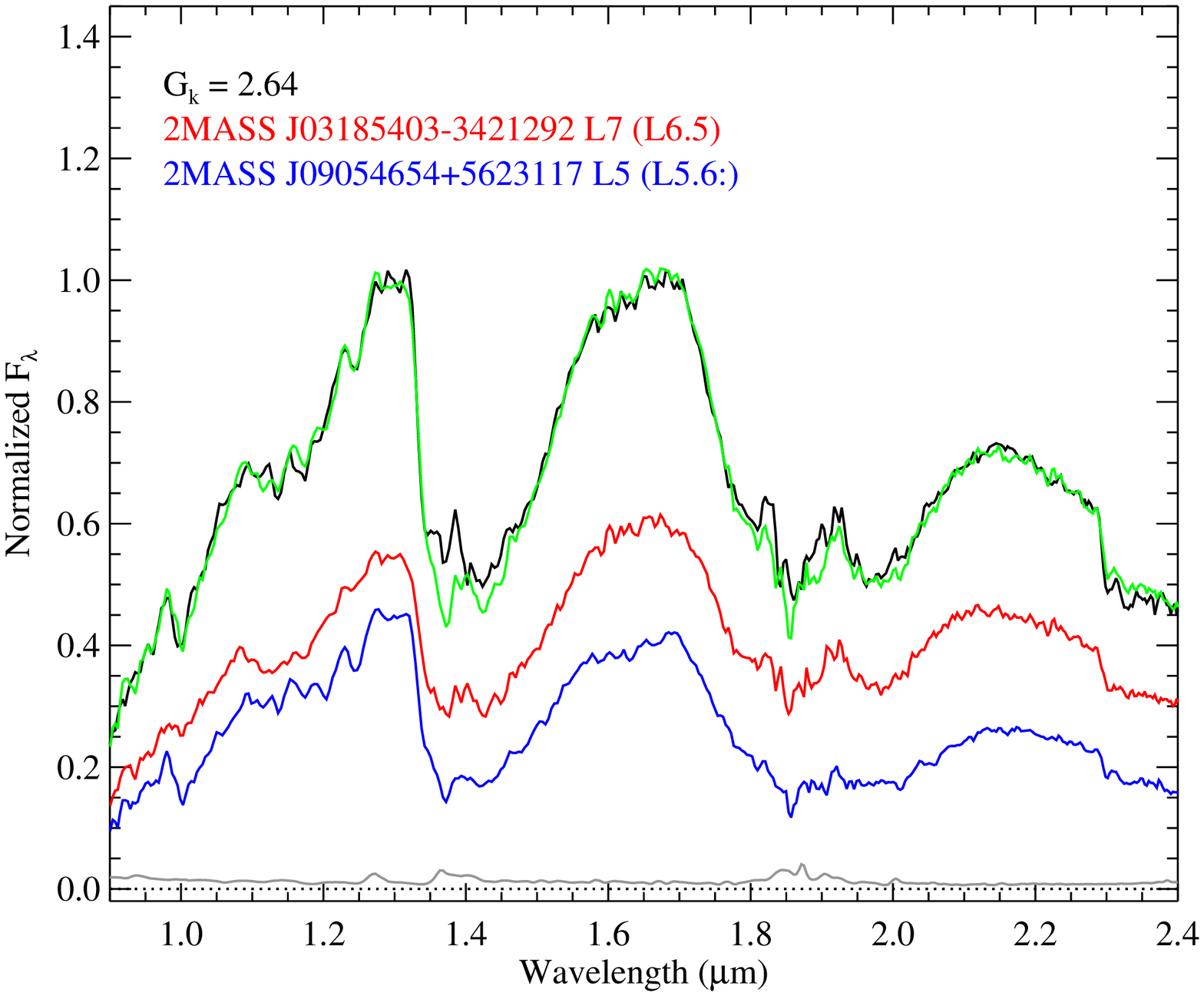}
\includegraphics[width=0.45\textwidth]{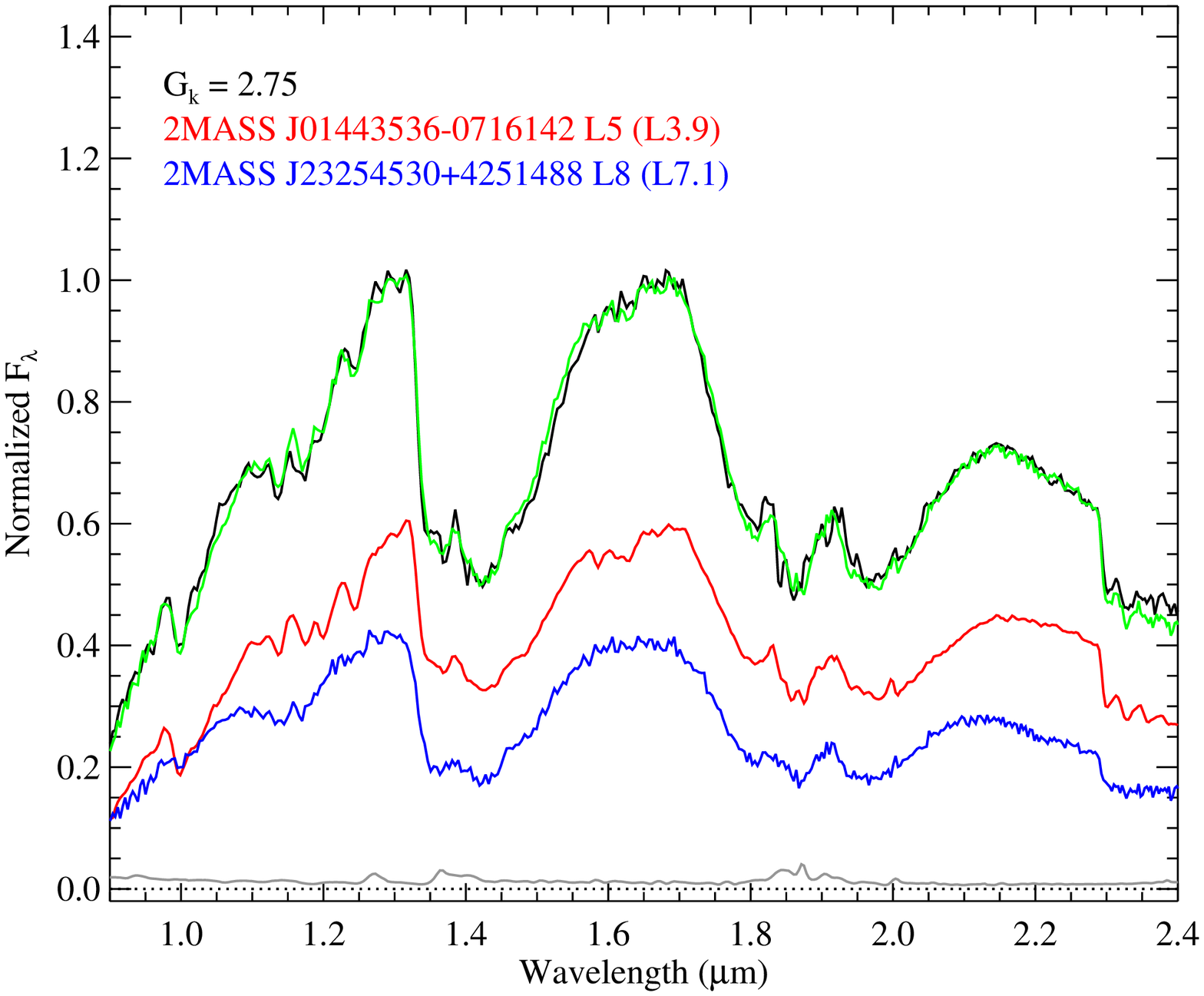}
\caption{The four best template fits for {\nameshb}, displayed as in Figure~\ref{fig_fit0850}.
Note that the L7 literature spectral type for SDSS~J104409.43+042937.6 is based on near-infrared data; all other literature types are based on optical data.
\label{fig_fit1728}}
\end{figure*}

The resulting best-fitting binary templates are listed in Table~\ref{tab_fits_all}, while Figures~\ref{fig_fit0850} and~\ref{fig_fit1728} display the four best fits to {\namesha} and {\nameshb}, respectively. 
These templates reproduce the observed data rather well, including the overall spectral slopes, shapes of the $JHK$ flux peaks, and depths of the {\wat}, CO and FeH absorption bands.  By design, each of the spectral combinations listed also reproduce the relative magnitude measurements to within the observational uncertainties.  Notably, the binary template fits are statistically superior to equivalent single template comparisons.  Best-fit $G_k$ values for these single templates were 2.20 for {\namesha} (comparing to the L5/L6.5\footnote{Hereafter, we list both literature and SpeX spectral types for each template; uncertain types are indicated with ``:'' for $\pm$1 subtype, ``::'' for $\pm$2 subtypes.} 2MASSW J1326201-272937; \citealt{2002ApJ...575..484G}) and 3.90 for {\nameshb} (comparing to the L5:/L5 2MASS J06244595-4521548; \citealt{2008AJ....136.1290R}), differing at the $>$ 99.9\% significance level based on the F-test statistic.  The significantly improved fits by the binary templates indicate that they are (as expected) better representations of the combined light spectra of these sources.

\begin{deluxetable*}{llllllccc}
\tabletypesize{\footnotesize}
\tablecaption{Best Fitting Binary Templates\label{tab_fits_all}}
\tablewidth{0pt}
%\rotate
\tablehead{
\colhead{Primary} &
\colhead{SpT} &
\colhead{SpT} &
\colhead{Secondary} &
\colhead{SpT} &
\colhead{SpT} &
\colhead{$G_k$} &
\colhead{Relative}  \\
 & \colhead{(Lit.)\tablenotemark{a}} & \colhead{(SpeX)\tablenotemark{b}}  &  & \colhead{(Lit.)\tablenotemark{a}} & \colhead{(SpeX)\tablenotemark{b}}  &  & \colhead{Weight\tablenotemark{c}} \\
}
\startdata
\multicolumn{8}{c}{\namea} \\
\cline{1-8}
SDSS J151506.11+443648.3 & L7.5$\pm$1.5\tablenotemark{d} & L6.9 & 2MASSW J1553214+210907 & L5.5 & L6.1: & 1.46 & 1.00 \\
SDSS J151506.11+443648.3 & L7.5$\pm$1.5\tablenotemark{d} & L6.9 & 2MASSI J0028394+150141 & L4.5 & L5.2:: & 1.57 & 0.28 \\
SDSS J151506.11+443648.3 & L7.5$\pm$1.5\tablenotemark{d} & L6.9 & SDSSp J010752.33+004156.1 & L8 & L6.0 & 1.62 & 0.20 \\
SDSS J171714.10+652622.2 & L4 & L5.8: & SDSSp J010752.33+004156.1 & L8 & L6.0 & 1.85 & 0.03 \\
SDSS J171714.10+652622.2 & L4 & L5.8: & SDSS J080959.01+443422.2 & L6\tablenotemark{d} & L6.2: & 1.88 & 0.02 \\
SDSS J171714.10+652622.2 & L4 & L5.8: & SDSSp J132629.82-003831.5 & L8: & L6.1: & 1.94 & 0.01 \\
2MASS J03101401-2756452 & L5 & L6.0: & SDSSp J010752.33+004156.1 & L8 & L6.0 & 1.95 & 0.01 \\
\multicolumn{1}{c}{\bf Weighted Mean} & L7.5$\pm$0.0 & L6.9$\pm$0.0 & \multicolumn{1}{c}{\bf Weighted Mean}  & L5.7$\pm$1.2 & L5.8$\pm$0.4 & \nodata & \nodata \\
\cline{1-8}
\multicolumn{8}{c}{\nameb} \\
\cline{1-8}
2MASS J01443536-0716142 & L5 & L3.9 & SDSS J104409.43+042937.6 & L7\tablenotemark{d} & L7.1 & 2.41 & 1.00 \\
2MASSW J2224438-015852 & L4.5 & L4.0: & SDSS J104409.43+042937.6 & L7\tablenotemark{d} & L7.1 & 2.58 & 0.29 \\
2MASS J03185403-3421292 & L7 & L6.5 & 2MASS J09054654+5623117 & L5 & L5.6: & 2.64 & 0.23 \\
2MASS J01443536-0716142 & L5 & L3.9 & 2MASS J23254530+4251488 & L8 & L7.1 & 2.75 & 0.15 \\
2MASS J03185403-3421292 & L7 & L6.5 & 2MASSI J1305410+204639 & L4: & L6.0: & 2.92 & 0.06 \\
2MASS J01443536-0716142 & L5 & L3.9 & SDSSp J003259.36+141036.6 & L8\tablenotemark{d} & L7.9: & 3.12 & 0.02 \\
\multicolumn{1}{c}{\bf Weighted Mean} & L5.3$\pm$1.0 & L4.6$\pm$1.1 & \multicolumn{1}{c}{\bf Weighted Mean}  & L6.5$\pm$1.1 & L6.8$\pm$0.6 & \nodata & \nodata \\
\enddata
\tablenotetext{a}{Optical spectral type from the literature unless otherwise noted.}
\tablenotetext{b}{Classifications based on {\wat} and {\meth} index/spectral type relations 
defined in \citet{2007ApJ...659..655B}, following the iterative procedure detailed in \citet{2010ApJ...710.1142B}.}
\tablenotetext{c}{Statistical weight assigned to template parameters in calculation of means and uncertainties.  This weight corresponds to the F-test probability distribution function
for the ratio $G_k$/min($G_k$) and effective degrees of freedom $\nu$ = 211 (see Equation~4 in \citealt{2010ApJ...710.1142B}).}
\tablenotetext{d}{Near-infrared classification from the literature.} 
\end{deluxetable*}

The spectra of the best-fit template components also replicate photometric trends from HST and LGSAO imaging. For {\namesha}, each of the templates in Figure~\ref{fig_fit0850} has a secondary that is distinctly redder than its primary.   These secondaries ---
2MASSW J1553214+210907 (L5.5/L6.1:, $J-K_s$ = 2.03$\pm$0.19; \citealt{1999ApJ...519..802K}),
2MASSI J0028394+150141 (L4.5/L5.2::, $J-K_s$ = 1.95$\pm$0.13; \citealt{2000AJ....120..447K}) and
SDSSp J010752.33+004156.1 (L8/L6.0, $J-K_s$ = 2.12$\pm$0.07; \citealt{2002ApJ...564..466G}) 
are among the reddest L dwarfs currently known.  In contrast, the 
primaries of these templates have $J-K_s$ colors that are in line with median values for their spectral types 
(e.g., \citealt{2010AJ....139.1808S}).  What is more surprising, however, is that the best-fit secondaries are generally of {\it earlier type} than the best-fit primaries.  This is true for both literature and SpeX classifications.
The mean component classifications reflect this: L7.5 and L5.5$\pm$1.2 based on literature classifications and L7 and L6 based on SpeX classifications; we adopt the latter combination as the uncertainty-weighted means.  This apparent reversal in classifications is particularly remarkable given the 0.8--1.2~magnitude brightness difference between the primary and secondary.  There is no evidence that either component is a T dwarf.

For {\nameshb}, the best-fit secondaries are consistently bluer than the primaries,
again in line with photometric trends.  Notably, the components of the binaries shown in Figure~\ref{fig_fit1728} have comparable fluxes in the 0.9--1.0~$\micron$ region, supporting the F1042M flux reversal reported by \citet{2003AJ....125.3302G}, and diverge toward longer wavelengths.   For these templates, it is the primary components that are unusually red:
2MASS J01443536-0716142 (L5/L4.2, $J-K_s$ = 1.92$\pm$0.03; \citealt{2003AJ....125..343L}),
2MASSW J2224438-015852 (L4.5/L4.5:,  $J-K_s$ = 2.05$\pm$0.04; \citealt{2000AJ....120..447K}) and
2MASS J03185403-3421292 (L7/L6.4, $J-K_s$ = 2.06$\pm$0.07; \citealt{2008ApJ...689.1295K}).
The mean spectral type of {\nameshb}A, L5.5$\pm$1.0 from literature classifications and L4.5$\pm$1.1 from SpeX classifications, is also remarkable for being considerably earlier than the combined-light L7 optical classification.  It is typically the earlier-type component that dominates optical flux in L and T dwarf binaries. In this case, it appears that the comparable brightnesses of the two components at red optical wavelengths produces a ``blended'' combined-light spectral type.
The secondaries of the best-fit templates have normal colors, although both 
SDSS J104409.43+042937.6 (L7/L6.9; \citealt{2004AJ....127.3553K})
and 2MASS J23254530+4251488 (L8/L7.3:; \citealt{2007AJ....133..439C}) 
show weak signatures of 1.6 and 2.2~$\micron$ {\meth} absorption in their near-infrared spectra.  
These features suggest that {\nameshb}B is 
at the threshold of the L dwarf/T dwarf transition, despite its mid-L near-infrared spectral type.
We adopt mean component types of L5 and L6.5 for this system.

\subsection{Assessment of Systematic Effects}

To assess the robustness of our results, we conducted the same template fits for three subsets of relative photometry: the NICMOS data alone, and single filter scalings with F110W and $K_p$ photometry.  The NICMOS fits produce little change in the inferred spectral components, with {\namesha} still hosting a later-type primary (L7 + L6 components based on both literature and SpeX classifications) and {\nameshb} having a primary that is typed earlier than the combined light spectrum (L5 + L6.5 for literature classifications, L4 + L7 for SpeX classifications).  However, the mean relative $K_p$ magnitudes from these fits --- 0.950$\pm$0.015 for {\namesha} and 0.54$\pm$0.05 for {\nameshb} --- differ from the measurements of \citet{2010ApJ...711.1087K} at the 7$\sigma$ and 1.5$\sigma$ levels, respectively.  Inferred brightnesses diverge even more dramatically when a single filter is used to scale the templates.  F110W-scaled templates for {\namesha} indicate component types of L6 + L9:, with a significantly reduced best-fit $G_k$ value (1.12 versus 1.46, distinct at the 98\% confidence level).  These types are more in line with prior estimates (e.g., \citealt{2001AJ....121..489R}), but the inferred component magnitudes differ significantly from measured values: 
$\Delta$F160W  disagrees at the 8$\sigma$ level, while $\Delta{K_p}$ disagrees at the 6.5$\sigma$ level. 
$K_p$-scaled templates for {\namesha} produce comparably large discrepancies in inferred NICMOS photometry.
Single-filter template fits to {\nameshb} are somewhat more robust, producing similar component types as the baseline template sample (L5+L7 for F110W fits, L5+L7.5: for $K_p$ fits), although inferred relative magnitudes still deviate by up to 2$\sigma$.

We conclude that multi-band photometry is essential to accurately and precisely constrain the component properties of these binary systems, and particular care must be taken when drawing conclusions from single-filter measurements.  Nevertheless, the general agreement in component types inferred from different multi-band subsets (e.g., NICMOS+$K_p$, NICMOS only) indicates that our results are robust.  
%We list our final component classifications (a weighted mean of literature- and SpeX-based classifications) in Table~\ref{tab_components}.

\section{Absolute Magnitude/Spectral Types and Color Trends}

\begin{figure*}
\centering
\epsscale{1.0}
\includegraphics[width=0.45\textwidth]{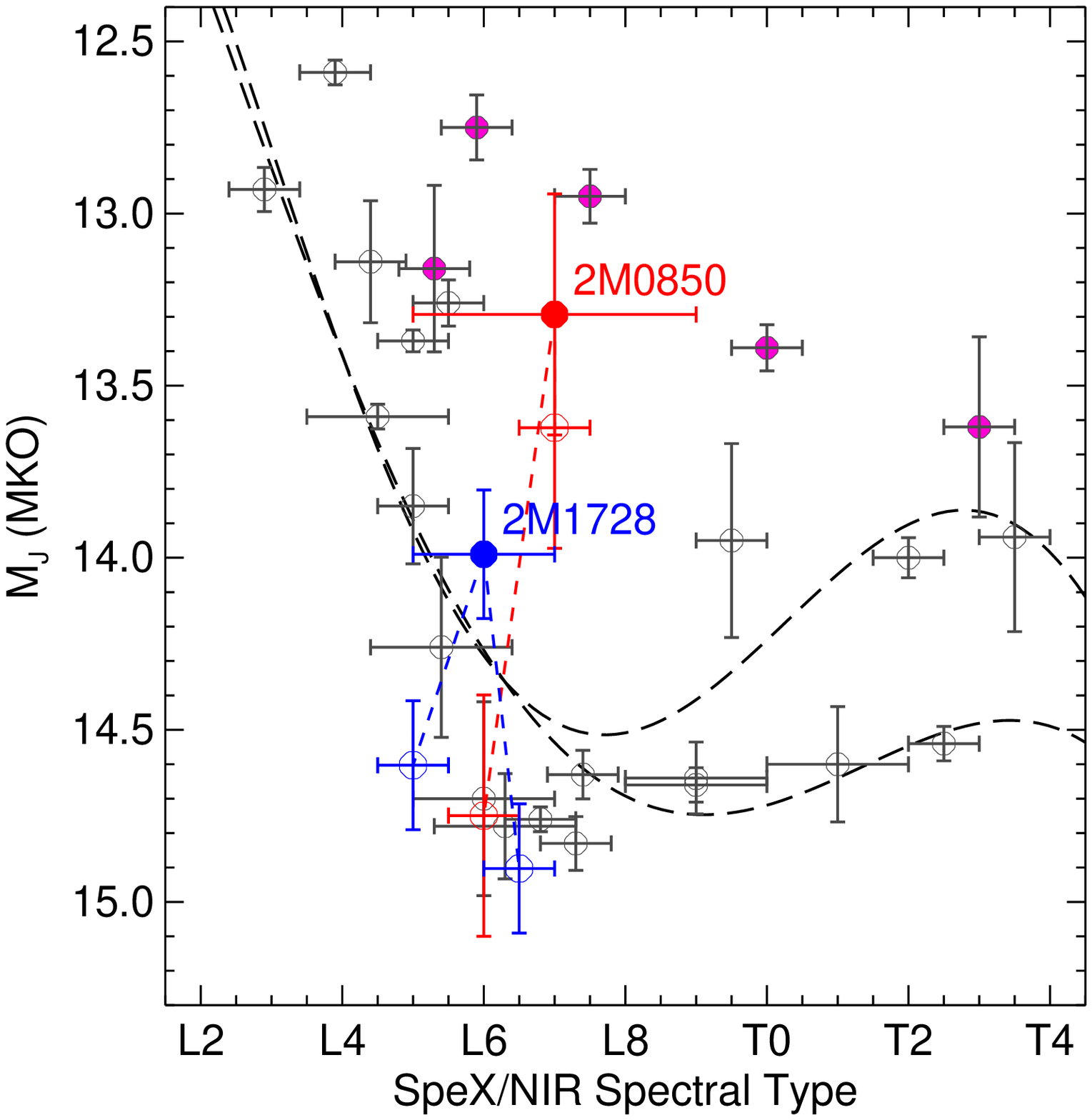}
\includegraphics[width=0.45\textwidth]{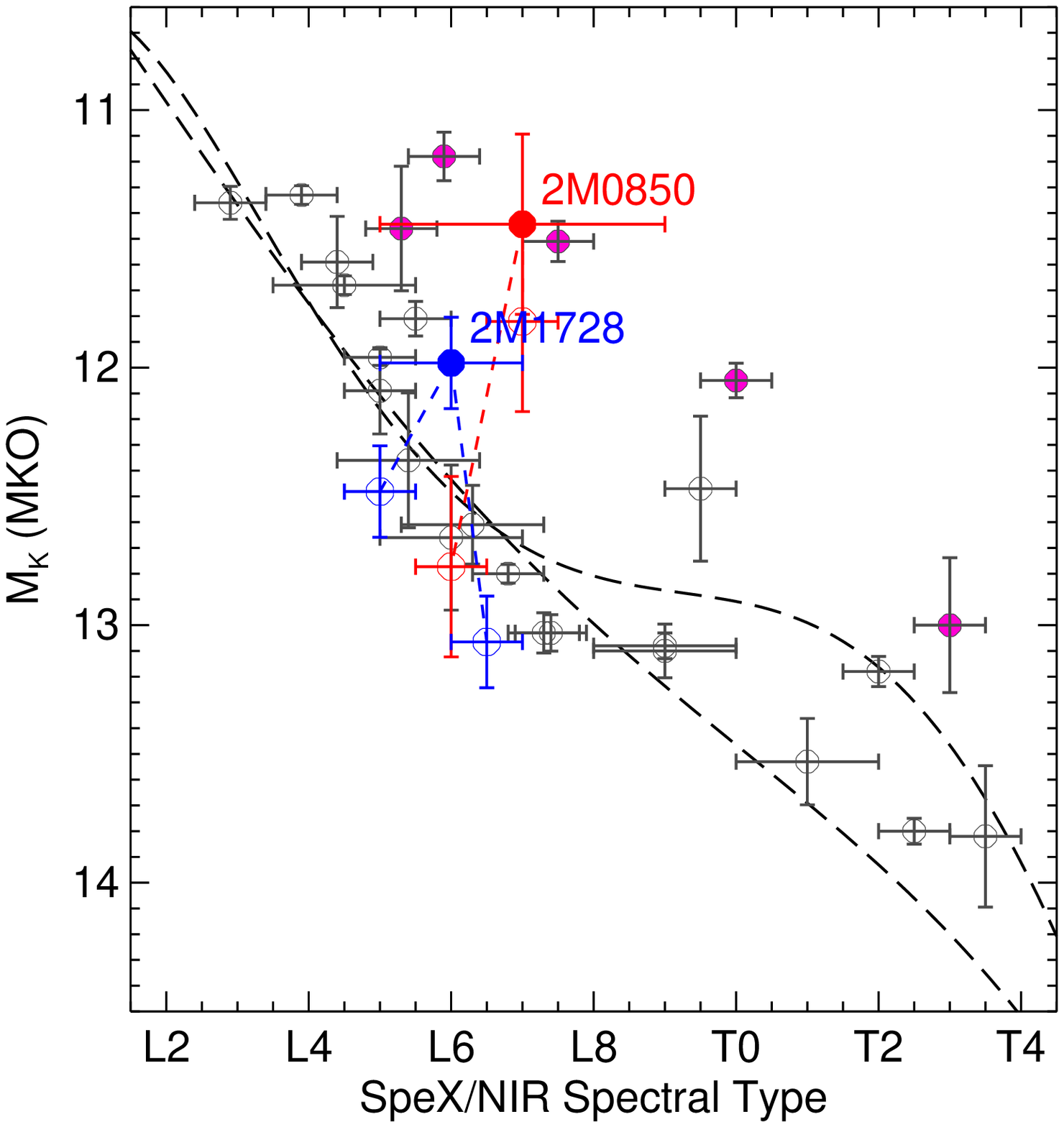}
\includegraphics[width=0.45\textwidth]{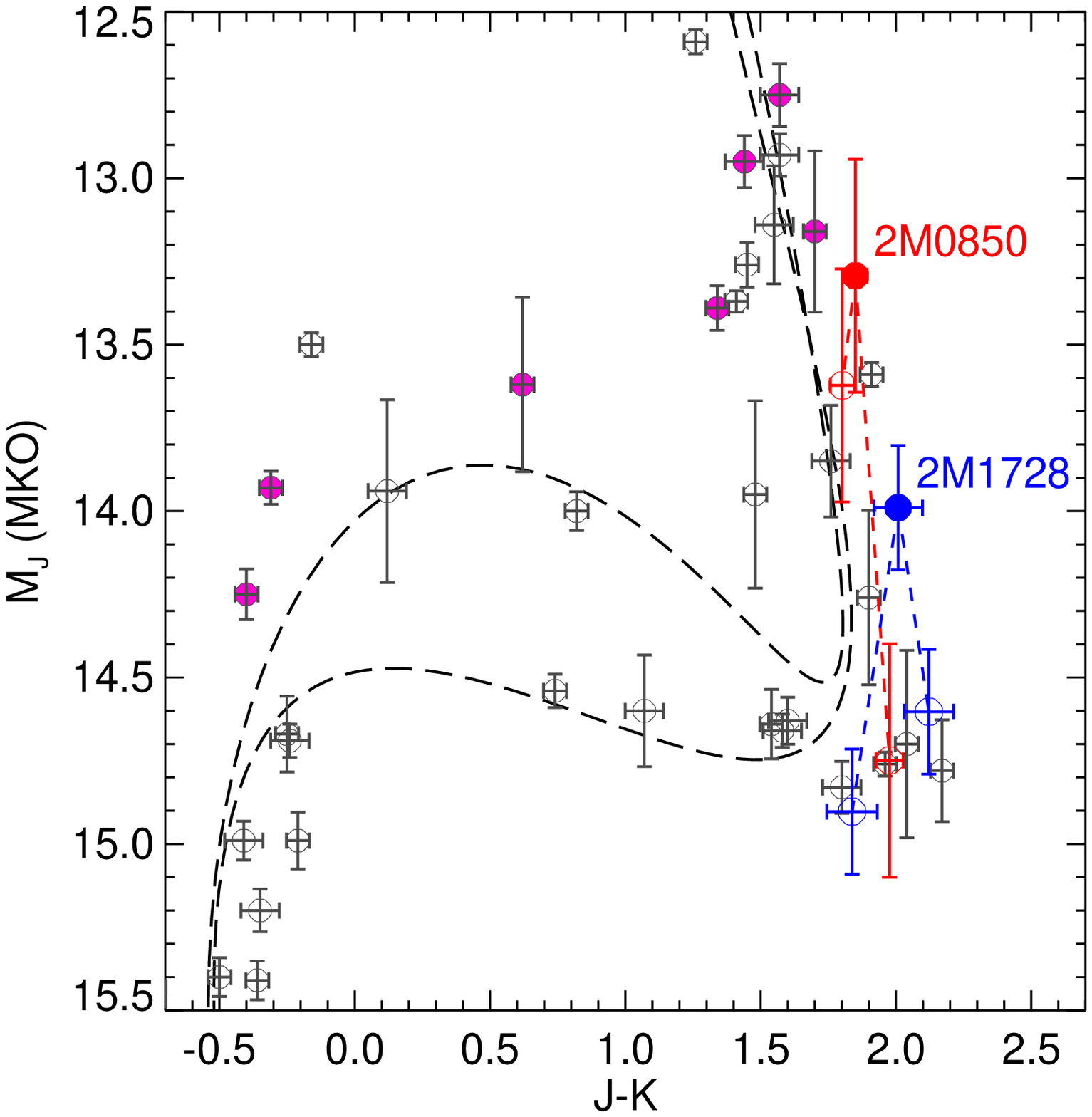}
\includegraphics[width=0.45\textwidth]{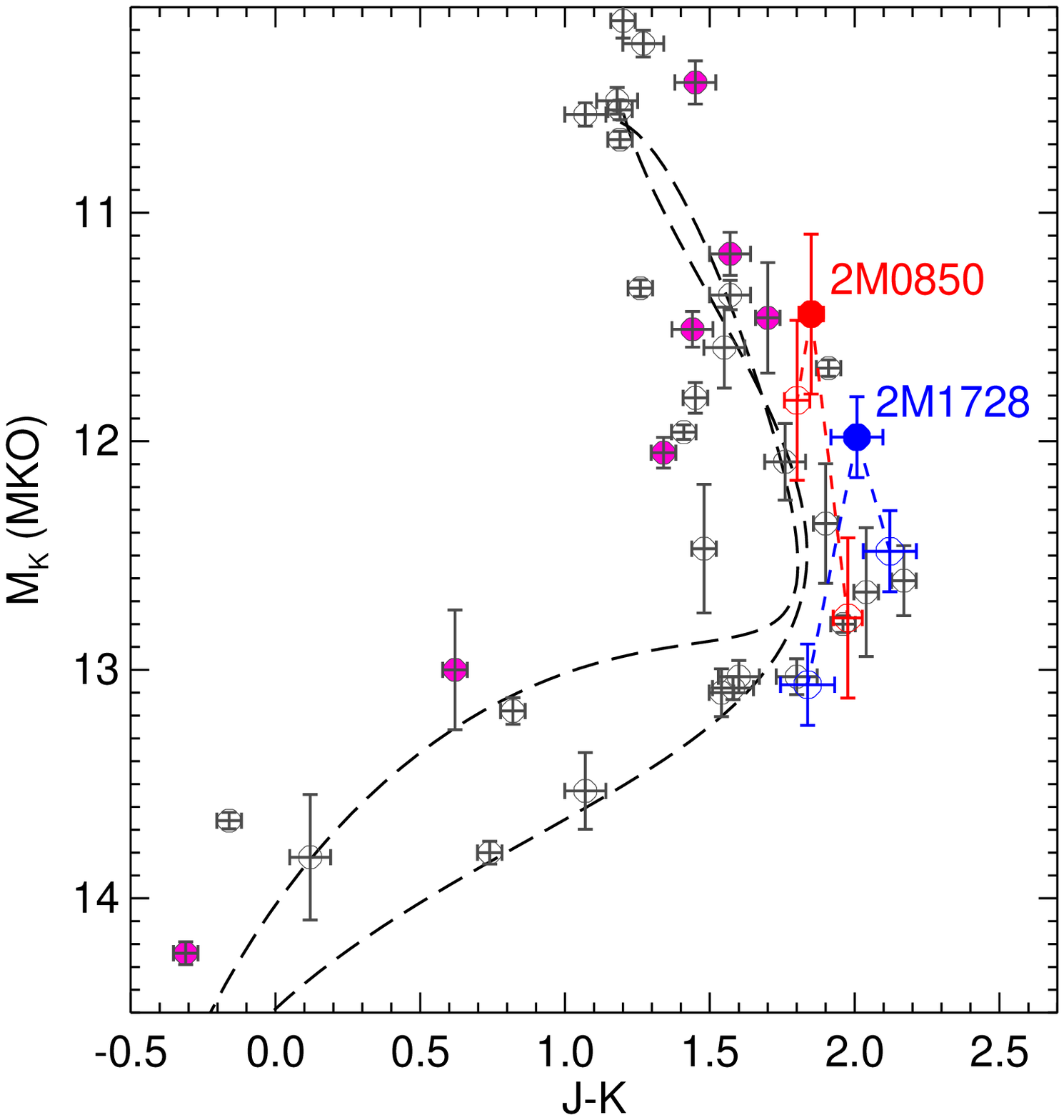}
\caption{(Top panels) Absolute MKO $J$ and $K$ magnitudes versus spectral type for the {\namesha} (red) and {\nameshb} (blue) components.  These are compared to single (open symbols) and binary L and T dwarfs (filled symbols) with absolute magnitude uncertainties $\leq$ 0.3~mag, as compiled by \citet{2010ApJ...710.1627L}.  All sources are plotted according to their near-infrared classifications, either published or computed from SpeX spectroscopy.  Also shown are the absolute magnitude/spectral type relations quantified in \citet[dashed lines]{2006ApJ...647.1393L}; both ``bright'' and ``faint'' relations are shown. Combined light photometry for 
{\namesha} and {\nameshb} are connected to component values. 
(Bottom panels) Absolute magnitudes versus MKO $J-K$ color based on the same sample.
\label{fig_abs}}
\end{figure*}

Both {\namesha} and {\nameshb} have parallax distance measurements from \citet{2004AJ....127.2948V}, so it is possible to 
compare component absolute fluxes to those of comparable field dwarfs.  Figure~\ref{fig_abs} displays combined light and component absolute MKO $J$ and $K$ magnitudes versus near-infrared spectral type and $J-K$ color.  The MKO magnitudes from {\namesha} and {\nameshb} are based on measurements from \citet{2002ApJ...564..452L} for the former, and synthesized from 2MASS photometry and SpeX spectroscopy for the latter.  Component $JK$ fluxes were determined from the template fits, synthesized from the various best-fit template spectra and combined following the same weighting scheme as the average spectral types (Table~\ref{tab_components}).  We use the near-infrared spectral types calculated in Section~2.2.
The comparison sample was drawn from the compilation of \citet{2010ApJ...710.1627L}, where we have either used published near-infrared spectral types listed in that study or types calculated from SpeX spectra (where available) following the same spectral index method described above.

\clearpage
\begin{deluxetable*}{lccccc}
\tabletypesize{\footnotesize}
\tablecaption{Adopted Component Parameters for {\namea} and {\nameb}\label{tab_components}}
\tablewidth{0pt}
\tablehead{
\colhead{Parameter} &
\multicolumn{3}{c}{\namesha} & 
\multicolumn{2}{c}{\nameshb} \\
 &
\colhead{A} &
\colhead{AA\tablenotemark{a}} &
\colhead{B} &
\colhead{A} &
\colhead{B} \\
}
\startdata
NIR Spectral Type & L7 & L7+L7 & L6 & L5 & L6.5 \\
$M_J$ & 13.62$\pm$0.35 & 14.37$\pm$0.35 & 14.75$\pm$0.35 & 14.60$\pm$0.19 & 14.90$\pm$0.19 \\
$M_K$ & 11.82$\pm$0.35 & 12.57$\pm$0.35 & 12.77$\pm$0.35 & 12.48$\pm$0.18 & 13.07$\pm$0.18 \\
 $J-K$ & 1.80$\pm$0.04 & \nodata & 1.98$\pm$0.05 & 2.12$\pm$0.09 & 1.84$\pm$0.09 \\
{\lbol}\tablenotemark{b} & -4.13$\pm$0.14 & -4.43$\pm$0.14 & -4.52$\pm$0.14 & -4.42$\pm$0.08 & -4.63$\pm$0.08 \\
{\teff} (K)\tablenotemark{c} & 1770$\pm$200 & 1480$\pm$180 & 1400$\pm$170 & 1570$\pm$80 & 1340$\pm$70 \\
Mass ({\msun})\tablenotemark{c} & 0.06$\pm$0.02 & 0.10$\pm$0.03 & 0.05$\pm$0.02 & 0.075$\pm$0.007 & 0.066$\pm$0.008 \\
\enddata
\tablenotetext{a}{{}Parameters assuming two equal-mass components. Mass is given for combined pair.}
\tablenotetext{b}{{}Luminosity based on $M_K$ magnitudes and $K$-band bolometric 
corrections ($BC_K$) from \citet{2009ApJ...704.1519D}. Stated 1$\sigma$ errors include $\pm$0.5 uncertainty in spectral type, absolute magnitude uncertainties, and 0.08~dex scatter in the Dupuy \& Liu $BC_K$/spectral type relation.}
\tablenotetext{c}{{\teff} and mass estimates based on inferred luminosities and ages listed in Table~\ref{tab_properties} ({\nameshb} is assumed to be 1.5--8~Gyr), combined with cloudy evolutionary models from \citet{2008ApJ...689.1327S}. Stated 1$\sigma$ errors include uncertainties in luminosities and spread in age estimates.}
\end{deluxetable*}

Focusing first on the absolute magnitude/spectral type comparisons, we find that three of the components---{\namesha}B, {\nameshb}A and {\nameshb}B---roughly cluster, with absolute magnitudes similar to other L6--L8 sources.  {\nameshb}A, and to a lesser degree {\nameshb}B, are somewhat underluminous for their types, particularly at $J$-band where they fall $\approx$0.5~mag below the absolute magnitude relations of \citet{2006ApJ...647.1393L}. 
%{\nameshb}B in particular appears to be the faintest L dwarf at $J$-band with a measured parallax. 
In contrast, the primary of {\namesha} is $\approx$1~mag brighter than comparable L6--L8 dwarfs,
or equivalently has a classification 3 subtypes too late for its measured absolute brightness.  This component is  even marginally brighter than combined light photometry for {\nameshb}.
As such, the discrepancy between brightness and spectral type between the components of the {\namesha} system appears to be rooted to the unusually late spectral type and/or overluminosity of its primary.

In the near-infrared color-magnitude diagrams (CMDs), all four components follow the locus of field L dwarfs, tracing the inflection in $J-K$ color and $M_J$ magnitude at the end of the L sequence.
In particular, {\namesha}B and {\nameshb}B  bridge the turnover to bluer near-infrared colors, and the latter appears to be the faintest L dwarf with a parallax measurement currently known.  Its location on the near-infrared CMDs supports our hypothesis that it is on the cusp of becoming a T dwarf (Section~4.2).
The components of {\namesha} are no longer outliers in these plots, in contrast to prior results (e.g., \citealt{2004AJ....127.2948V}). {\namesha}A in particular does not stand out as unusually bright, due largely to the near-vertical locus of L dwarfs in near-infrared CMDs.

For completeness, we also determined luminosity, {\teff} and mass estimates for the components of these binaries using their inferred absolute magnitudes and near-infrared spectral types.  Bolometric luminosities were calculated using $K$-band bolometric corrections ($BC_K$) derived from the $BC_K$/spectral type relation of \citet{2009ApJ...704.1519D}.  Note that these values may be systematically too high for the unusually red sources {\namesha}B and {\nameshb}A.  {\teff}s and masses were estimated using the \citet{2008ApJ...689.1327S} cloudy evolutionary models, based on the derived luminosities and age estimates listed in Table~\ref{tab_properties}; for {\nameshb}, we assumed an age range of 1.5--7.5~Gyr, the upper limit based on the small {\vtan} of this source.  The inferred {\teff}s for {\nameshb}A, {\namesha}B and {\nameshb}B are consistent with prior estimates for L5--L6.5 dwarfs \citep{2009ApJ...702..154S}, while the brightness of {\namesha}A makes it $\approx$300~K warmer than comparable L7 dwarfs.  We discuss this component in detail in Section~6.2.

\section{Discussion}

\subsection{Clouds and Classification}

Our photometric and spectroscopic analyses of the components of {\namesha} and {\nameshb} have revealed several unusual traits, particularly in their primaries: under- and overluminous fluxes and unusually early and late near-infrared spectral types.  These peculiar traits can be related to their unique atmospheric properties.  

In the case of {\nameshb}A, we hypothesize that condensate cloud effects are responsible for shifting this component toward both an earlier near-infrared spectral classification and toward slightly fainter $J$-band fluxes.
Grey extinction from condensate cloud grains in L dwarf photospheres dominate the opacity at the $J$- and $H$-band flux peaks, as these windows in molecular gas opacity probe deeper into the atmosphere and sample a larger column depth of cloud material \citep{2001ApJ...556..872A, 2006ApJ...640.1063B}. The $K$-band peak, on the other hand, is modulated by both cloud opacity and collision-induced $H_2$ absorption \citep{1969ApJ...156..989L}.  As a result, greater condensate opacity tends to produce redder $J-K$ colors and fainter $J$-band fluxes \citep{2001ApJ...556..357A, 2002ApJ...568..335M, 2004ApJ...607..511T}.  In addition, contrast in molecular absorption bands is reduced, particularly for the near-infrared {\wat} and FeH bands to which near-infrared schemes are commonly tied \citep{2001AJ....121.1710R,2002ApJ...564..466G}. Veiling of these features can skew near-infrared classifications toward earlier types \citep{2003IAUS..211..355S, 2004AJ....127.3553K}.  
Hence, thicker condensate clouds leads to redder near-infrared colors, reduced flux at $J$-band, and systematically earlier near-infrared classifications.

These conditions accurately reflect the properties of {\nameshb}A, and to a lesser degree {\namesha}B.  The best-fit primary components for {\nameshb} are consistently red, mid-type L dwarfs whose near-infrared spectral types are consistently earlier than their optical types (Table~\ref{tab_fits_all}).  These include the templates
2MASS J01443536-0716142 (hereafter 2MASS~J0144-0716; L5 optical classification, L4 SpeX classification; \citealt{2003AJ....125..343L}) and
2MASSW J2224438-015852 (hereafter 2MASS~J2224-0158; L4.5 optical classification, L3.5 near-infrared classification; \citealt{2000AJ....120..447K,2004AJ....127.3553K}).
%2MASSW J0030300-145033 (hereafter 2MASS~J0030-1450; L7 optical classification, L4.5 near-infrared classification; \citealt{2000AJ....120..447K}; this paper). 
Both of these sources exhibit indications of thick clouds, based on 
the detection of linear polarization in the case of 2MASS~J0144-0716\footnote{2MASS~J0144-0716 was also detected in an optical flare by \citet{2003AJ....125..343L}.} (0.58$\pm$0.19\% at $I$-band; \citealt{2005ApJ...621..445Z}), and spectral model fits and pronounced silicate grain absorption at 9--11~$\micron$ in the case of 2MASS~J2224-0158 \citep{2006ApJ...648..614C, 2008ApJ...678.1372C,2009ApJ...702..154S}.
% and evidence of photometric variability in the case of 2MASS~J0030-1450 \citep[however, see \citealt{2008MNRAS.386.2009C}]{2003AJ....126.1006E}.
Unusually red dwarfs such as these have been found to have cooler {\teff}s for their
spectral types \citep{2009ApJ...702..154S}, which can contribute to fainter magnitudes.
The older age of the {\nameshb} system, based on the absence of {\lii} absorption, argues that thick condensate clouds, rather than low surface gravity, gives rise to its unusually red color \citep{2007ApJ...657..511A,2008ApJ...686..528L}.

Thick clouds in {\nameshb}A may also explain its comparable brightness at 1~$\micron$ compared to its later-type companion.  As discussed above, {\nameshb}B appears to be at the threshold of the L dwarf/T dwarf transition, a phase in which condensate clouds are inexplicably dispersed and $J$-band fluxes increase.  In a sample of unresolved L/T transition binary candidates, \citet{2010ApJ...710.1142B} found that sources with comparable component types but redder primaries showed a more pronounced flux reversal than sources with normal or blue primaries.  They argued that $J$-band fluxes in the primaries of these systems were more suppressed.  We may be seeing a similar effect in the {\nameshb}AB pair.
Alternatively, we may be observing the tops of thick clouds that are constrained to the same temperature layer in both sources (i.e., T$_{cr}$ = constant; \citealt{2005ApJ...621.1033T}).  These possibilities should be explored with 
detailed modeling of resolved component spectra, rather than spectral templates.

\subsection{Is {\namesha} a Young Triplet?}

While thick clouds can explain the unusual faintness of {\nameshb}A,
thin clouds do not readily explain the unusual brightness of
{\namesha}A.  Thin-cloud L dwarfs, also known as unusually blue L dwarfs (UBLs),
do exhibit discrepancies between optical and near-infrared spectral types \citep{2004AJ....127.3553K,2008ApJ...674..451B}. There is also evidence that UBLs
have systematically larger {\teff}s and/or absolute fluxes for their spectral classifications
\citep{2009ApJ...702..154S}.  However, {\namesha}A has a normal near-infrared color for its spectral type, and best-fit templates exhibit none of the spectroscopic hallmarks of a UBL (e.g., deep {\wat} and FeH bands, blue near-infrared SED; \citealt{2008ApJ...674..451B, 2010AJ....139.1045S}).
It is therefore unlikely that this source is a UBL with unusually thin clouds.

Youth may play a role in the unusual brightness of {\namesha}A, arising from the enlarged radius of a still-contracting brown dwarf.  The radius of a 1500~K brown dwarf is 25\% larger at an age of 250~Myr --- the minimum bound cited by \citet{faherty0850} ---  than at 2~Gyr, corresponding to an increase in brightness of roughly 0.5~mag.  Given the current uncertainties in the distance of this source, such a shift may be sufficient to move {\namesha}A back onto absolute magnitude/spectral type tracks.  However, assuming that the two components of this system are coeval, {\namesha}B would also have to be overluminous by roughly the same factor, which does not appear to be the case.  Moreover, such a correction fails to explain the significant brightness difference between these two comparably-classified sources.

We propose an alternative explanation: {\namesha}A is itself an
unresolved, near-equal mass binary.  Such a scenario would explain how
this component could appear both brighter and warmer than its equivalently-typed
companion, but have an otherwise normal spectral energy distribution.
The components of an equal-mass {\namesha}A pair would have absolute magnitudes, luminosities and {\teff}s fully consistent with empirical trends (Table~\ref{tab_components}).
The fact that existing high angular resolution images have not resolved this source requires 
an angular separation $\lesssim$50--100~mas, or a projected separation $\lesssim$2--4~pc at this distance of this system.  In fact, long-term dynamic stability requires an even tighter binary.
A hierarchical triple is generally stable if the ratio of outer periastron and inner apastron distances:
\begin{equation}
Y \equiv \frac{a_{out}(1-e_{out})}{a_{in}(1+e_{in})}
\end{equation}
satisfies
\begin{equation}
Y > Y^{min} \equiv 1 + \frac{3.7}{q_{out}^{1/3}} + \frac{2.2}{q_{out}^{1/3}+1}
 + \frac{1.4}{q_{in}^{1/3}}\frac{q_{out}^{1/3}-1}{q_{out}^{1/3}+1}.
\label{eqn_orbit}
\end{equation}
\citep{1995ApJ...455..640E}.
Here, $q_{in} \equiv \frac{M_1}{M_2}$ $\geq$ 1 and $q_{out} \equiv \frac{M_1+M_2}{M_3}$
are the inner and outer mass ratios.  Assuming that all three components have nearly equal masses, and that the inner and outer orbits have eccentricities $e$ = 0 (0.5), Equation~\ref{eqn_orbit} requires a limit on the inner orbit semimajor axis of $a_{in} < 0.2a_{out}$ ($<$0.07$a_{out}$) and hence $a_{in} <$ 25~mas (8~mas), or 1~AU (0.3~AU), based on the semimajor axis determination of \citet{2010ApJ...711.1087K} and the \citet{2004AJ....127.2948V} parallax.  
%The larger of these two values is smaller than the resolution limit of HST and current LGSAO instrumentation (50--100~mas), and is comparable to the parallax of this system.  The smaller of these limits is comparable to the difference between the Dahn et al.\ and Vrba et al.\ parallax measurements.  Because the orbital period of such a tight binary is similar to the timespan over which astrometric observations are typically made (3.1~yr in the case of $e = 0$, 0.6~yr in the case of $e=0.5$), the astrometric wobble induced by a third component may be sufficient to systematically bias parallax measurements.  On the other hand, the nearby background source identified by \citet{faherty0850} may be sufficient on its own to explain the discrepancies in parallax measurements for {\namesha}.

Such a tight separation is not unusual for brown dwarf
multiples; brown dwarf spectroscopic binaries with comparable separations have already been identified (e.g., \citealt{1999AJ....118.2460B,2008ApJ...678L.125B, 2010arXiv1006.2383J}).
Moreover, three other tight, hierarchical, brown dwarf triple candidates
identified in the literature---Gliese~569BCD, \citep{2006ApJ...644.1183S},
DENIS~J0205-1159ABC \citep{2005AJ....129..511B}, and
Kelu~1ABC \citep{2008arXiv0811.0556S}---exhibit evidence that
one component is an unresolved pair, based on radial velocity variations, PSF-fitting
residuals and spectroscopic features, respectively.
This scenario is also consistent with orbital mass constraints from \citet{2010ApJ...711.1087K}, as the estimated total mass of a triple {\namesha} system, 0.15$\pm$0.04~{\msun}, is closer to the mean (but weakly constrained) value of 0.2~{\msun} found in that study.
Indeed, tighter constraints on the total mass of this system from ongoing astrometric monitoring may affirm or refute the presence of a third body.
The triple hypothesis can also be tested 
though radial velocity monitoring; a $\lesssim$5~{\kms} ($\lesssim$8~{\kms}) line shift arising from a pair of 0.05~{\msun} brown dwarfs separated by 1.0~AU (0.3~AU) can be readily detected with current near-infrared instrumentation \citep{2007ApJ...666.1198B,2010ApJ...711L..19B}.

If {\namesha}A is confirmed as a binary, it would complete a remarkable, young, low-mass, hierarchical quintuple system with the double M dwarf NLTT~20346AB, encompassing 4 orders of magnitude in separation and composed entirely of objects less massive than 0.15~{\msun} \citep{faherty0850}.
 
\section{Summary}

We have presented photometric and spectroscopic analyses of the late-type L dwarf
binaries {\namesha} and {\nameshb}, aimed at assessing component spectral types,
absolute magnitudes and near-infrared colors.  Multi-band HST/NICMOS photometry have revealed distinct trends in the relative colors of these two systems, with {\namesha}B being redder than its primary and {\nameshb}B being bluer.  Neither secondary exhibits narrow-band colors consistent with being a T dwarf.  These results are borne out in spectral template fits, using NICMOS and $K_p$ resolved photometry, which also determine component near-infrared spectral types of L7 + L6 for {\namesha} and L5 + L6.5 for {\nameshb}.  The early classification of {\nameshb}A, its relative faintness at $J$, and its unusually red color can be explained by the presence of thick condensate clouds in its photosphere.  The secondary of this system, in contrast, may be losing its photospheric cloud deck as it transitions onto the T dwarf sequence.  For {\namesha}, the surprisingly later spectral type of its bright primary may stem from youth (inflated radius) and/or unresolved multiplicity.  The latter hypothesis, which would make {\namesha} part of a low-mass hierarchical quintuple, can be tested through ongoing astrometric monitoring and/or resolved spectroscopic monitoring to search for RV variations.

As two resolved (or partly-resolved) coeval systems spanning the end of the L dwarf sequence and exhibiting a broad range of cloud properties, {\namesha} and {\nameshb} remain important laboratories for studying cloud formation and evolution in low-temperature atmospheres.  Improved parallactic distance measurements---including resolution of current distance discrepancies for {\namesha}---resolved component spectroscopy, and ongoing photometric and astrometric monitoring will aid in characterizing the clouds, spectral properties and multiplicity of these benchmark brown dwarf systems.

\acknowledgments

The authors would like to thank telescope operators Dave Griep and Bill Grolisch
and instrument specialist John Rayner
for their assistance during the IRTF observations.
We acknowledge helpful comments from Trent Dupuy and Jacqueline Faherty on our original manuscript, and thank our referee Sandy Leggett for her prompt and helpful review.
AJB acknowledges support from the Chris and Warren Hellman Fellowship; DCBG acknowledges funding from the John Reed Fund.
This publication makes use of data 
from the Two Micron All Sky Survey, which is a
joint project of the University of Massachusetts and the Infrared
Processing and Analysis Center, and funded by the National
Aeronautics and Space Administration and the National Science Foundation.
2MASS data were obtained from the NASA/IPAC Infrared
Science Archive, which is operated by the Jet Propulsion
Laboratory, California Institute of Technology, under contract
with the National Aeronautics and Space Administration.
This research has also made use of the SIMBAD database,
operated at CDS, Strasbourg, France; the M, L, and T dwarf compendium housed at DwarfArchives.org and maintained by Chris Gelino, Davy Kirkpatrick, and Adam Burgasser; the SpeX Prism Spectral Libraries, maintained by Adam Burgasser at \url{http://www.browndwarfs.org/spexprism}; and the VLM Binaries Archive maintained by Nick Siegler at \url{http://www.vlmbinaries.org}. 
The authors wish to recognize and acknowledge the 
very significant cultural role and reverence that 
the summit of Mauna Kea has always had within the 
indigenous Hawaiian community.  We are most fortunate 
to have the opportunity to conduct observations from this mountain.

Facilities: \facility{IRTF (SpeX)}, \facility{Hubble Space Telescope (NICMOS)}

\clearpage

\clearpage

%\bibliography{../../biblibrary}

\end{document}